\documentclass[trans]{IEEEtran}
\usepackage{hyperref}
\usepackage{pgfplots}
\pgfplotsset{compat=newest}
\usepackage{lettrine}
\usepackage{epsfig}
\usepackage{footnote}
\usepackage{cite}
\usepackage{float}
\usepackage{amssymb}
\usepackage{amsthm}
\usepackage{amsmath,mathabx}
\usepackage{color}
\usepackage{amsfonts}
\usepackage{epstopdf}
\usepackage{tabularx,tabulary}
\usepackage[font=footnotesize]{caption}
\usepackage{graphicx}
\usepackage{caption}
\usepackage{subcaption}
\usepackage{wrapfig}
\usepackage{dblfloatfix}    % To enable figures at the bottom of page
\usepackage{bbm}
\usepackage{algorithm,algpseudocode}
\makeatletter
\newcommand{\removelatexerror}{\let\@latex@error\@gobble}
\def\BState{\State\hskip-\ALG@thistlm}
\makeatother

\newcommand{\vect}[1]{\boldsymbol{#1}}

\definecolor{violet}{rgb}{1.0, 0.49, 0.0}

\begin{document}
\title{LightFDG: An Integrated Approach to Flow Detection and Grooming in Optical Wireless DCNs}
%\title{Flow Classification and Grooming for Wireless DCNs: A Fast, Light, and Accurate Flow Detection}

\author{
Amer~AlGhadhban,~\IEEEmembership{Member,~IEEE,} Abdulkadir~Celik,~\IEEEmembership{Senior Member,~IEEE,}\\ Basem Shihada,~\IEEEmembership{Senior Member,~IEEE,} and Mohamed-Slim~Alouini,~\IEEEmembership{Fellow,~IEEE}.

\thanks{
The authors are with Computer, Electrical, and Mathematical Sciences and Engineering (CEMSE) Division at King Abdullah University of Science and Technology (KAUST), Thuwal, 23955-6900, KSA. Amer~AlGhadhban is also with Electrical Engineering Department at University of Hail, Hail, 55474, KSA. 
%The associate editor coordinating the review of this paper and recommending it for publication was C. Assi.
(Corresponding author: Abdulkadir Celik, email: abdulkadir.celik@kaust.edu.sa)
This article was presented in part at the IEEE GLOBECOM 2018, Abu Dhabi, UAE \cite{Celik2018LightFD}. 

%Digital Object Identifier 10.1109/TNSM.2019.XXXXXXX

}
}

\markboth{IEEE Transactions on Network and Service Management}{}
\maketitle
\begin{abstract}
LightFDG is an integrated approach to flow detection (FD) and flow grooming (FG) in optical wireless data center networks (DCNs), which is interconnected via wavelength division multiplexing (WDM) based free-space optical (FSO) links. Since forwarding bandwidth-hungry elephant flows (EFs) and delay-sensitive mice flows (MFs) on the same path can cause severe performance degradation, the LightFDG optically grooms flows of each class into rack-to-rack (R2R) flows. Then, R2R-MF and R2R-EF flows are separately forwarded over lightpaths of separate MF and EF virtual topologies, respectively. Lightpaths are provisioned by jointly determining the capacity and route based on flows' arrival rate, size, and completion time request. To prevent EFs from congesting the MF lightpaths, high speed and accurate flow-detection mechanisms are also necessary for classifying EFs as soon as possible. Therefore, a fast-lightweight-and-accurate flow detection framework is developed by leveraging the transmission control protocol (TCP) behaviors. The proposed FD scheme has the flexibility of being implemented as in-network or centralized to classify flows of modifiable and unmodifiable hosts, respectively. Since the centralized scheme incurs considerable overhead, the processing and communication overhead is also mitigated by proposed techniques. Numerical results show that LightFDG outperforms traditional load balancers by about 3$\times$ for EFs and 10$\times$ for MFs. Along with the developed overhead mitigation methods, the centralized scheme is shown to provide up to 62$\times$ lower overhead with 100\% accuracy and with about 224$\times$ higher detection speeds than the existing centralized solutions. 
\end{abstract}


\section{Introduction}
\subsection{Preliminaries}
%\lettrine{T}{he} 
The cloud service providers, mobile operators, electronic commerce companies, and content delivery services strive to enhance the performance of their data center networks (DCNs) for keeping their services up with the recent technological trends \cite{Celik2018LightFD}. These include popular bandwidth-hungry services such as the fifth generation (5G) networks and beyond, the Internet of things, smart cities, big data, artificial intelligence, etc., that continuously exchange huge volume of diverse and complex data. To store and process such a deluge of data, DCNs are required to interconnect a plethora of servers with high-speed low-latency links, which eventually causes a significant cost and complexity related to wire deployment, maintenance, and modifications \cite{WDCNsurvey}. 

\subsection{Problem Statement and Motivation}
Traditional wired DCN topologies with fixed and uniform link capacities have been shown to be deficient in supporting the optimal capacity allocation mechanisms for aforementioned applications \cite{flowchar, skandula, facebook, projector}. This is mainly because of the fact that the majority of inter-rack traffic is exchanged between few racks whereas remaining racks exchange less traffic or no traffic at all. For instance, Hadoop servers of Facebook DCNs~\cite{facebook} communicate to only 1.5\% of the other inter-rack servers, which comprises only 17\% of the racks while approximately contributing to the 85\% of entire DCN traffic. This pattern is aligned with previous findings in \cite{skandula} where inter-rack traffic is carried out by up to 10\% of the servers. Thereby, the DCN topologies with uniform link capacities are either under-utilized compared to the rack pairs that exchange less traffic or over-utilized in comparison with the rack pairs that generate larger traffic sets \cite{fattree, vl2}, which is also known as \textit{over-subscription}.  

As a remedy wireless DCNs (WDCNs) can provide necessary reconfigurability to alleviate over-subscription, deliver a higher throughput thanks to efficient bandwidth utilization, eliminate the need for cost and power inefficient switches, and reduce the cabling cost \& complexity \cite{Celik2018magWDCN}. Required high-speed wireless links can be obtained by state-of-the-art technologies such as millimeter wave \cite{zhou2012mirror} and free-space optical (FSO) communications \cite{projector, Hamza17}. In particular, extremely high link capacities of FSO links can be further improved by wavelength division multiplexing (WDM) methods which yield a high fan-out (i.e., large number of channels per FSO link) as well as elastic topologies \cite{Hamza18}. Recently, Ciaramella et al.~\cite{Ciaramella2009128} achieved a total of 1.28 Tbps speed in an outdoor experiment over 212 meters distance by using WDM-FSO link of 32 wavelengths (32$\times$40 Gbps). Thanks to the acclimatized environment of DCNs, indoor FSO links can attain even better performances at longer distances since they are less vulnerable to the optical channel impairments such as scintillation, alignment errors, atmospheric turbulence, etc. 

In this paper, a spine-leaf WDCN architecture is considered which is inter-connected via WDM-FSO links. Since bandwidth demands of flows can be much lower than the available capacity on WDM channels, aggregating (i.e., grooming) sub-wavelength flows onto high-speed lightpaths can offer desirable link utilizations \cite{Celik2019Optical,Hamza19}. Considering the enormous amount of flows generated across the DCNs, flow grooming (FG) is also useful for avoiding unnecessary topology reconfiguration delays caused by the computational and control overhead of handling flows individually. Therefore, merging FG with the flexibility of high fan-out WDM-FSO links may provide necessary performance enhancement required by DCNs.

QoS provisioning is another crucial task to meet the diversified demands of different flow classes. It has been shown that the majority of bytes are carried by bandwidth-hungry elephant flows (EFs), while the majority of flows are delay-sensitive mice flows (MFs) \cite{flowchar, skandula, facebook, projector}. To reap the real benefits of reconfigurability, the FG policy should also adapt the network topology to provision the diverging QoS requirements of varying flow classes. From a load balancing point of view, MFs experience a severe delay if they are routed along the same path of EFs~\cite{hermes}. Therefore, grooming and forwarding the same class of flows altogether is a promising solution to handle heavy load and highly dynamic traffic characteristics of DCNs \cite{Celik2018Design, Celik2019Design}. In this regard, fast and precise flow classification schemes are necessary to groom the same class of flows and then steer them to their dedicated virtual-topology. Unfortunately, existing flow classification mechanisms have major drawbacks in terms of detection speed, controlling overhead, and accuracy.

Moreover, a DCN is a heterogeneous environment where the hosts can be modifiable or unmodifiable as per the DC administrator's right and capability to make changes \cite{clove}. In the former, system administrator can make modifications at user or kernel levels, which is subject to an agreement between the service provider and the subscriber. In the latter, the system modifications may not be possible for two reasons: 1) Changes are prohibited if a server and/or virtual machines (VMs) on the server belong to an identity other than the system administrator, and 2) Changes are not possible if a VM runs on a closed-source operating system (OS). Since the system administrator cannot modify all hosts, both centralized and in-network flow detection mechanisms might be needed.
 
Accordingly, this paper proposes LightFDG; an integrated flow detection, grooming, and forwarding approach. The LightFDG first determines the flow classes in a fast and accurate fashion, then grooms the same flow classes to obtain R2R flows, and finally forwards these aggregated flows over dedicated virtual topologies.

\subsection{Paper Objectives and Achievements}
Main contributions of this work can be summarized as follows:
 \begin{itemize} 
\item[$\bullet$] 
In order to avoid congestion and performance degradation, the proposed flow grooming and forwarding (FGF) policy is designed to handle the two main flow classes (i.e., MFs and EFs) by separately routing them  over disjoint virtual topologies. To overcome the complexity of handling each flow individually, a three-step flow grooming approach is adopted to obtain rack-to-rack (R2R) MFs and EFs. By jointly determining the route \& capacity of R2R MFs/EFs, the proposed FGF policy innately balances the traffic load across the WDM-FSO links and efficiently utilizes the available bandwidth while provisioning the QoS demands of each flow class. The emulation results show that the proposed FGF policy outperforms traditional load balancing schemes by about 3$\times$ for EFs and 10$\times$ for MFs.
 
\item[$\bullet$]
Instead of using traditional sampling methods, LightFDG employs a fast, lightweight, and accurate flow detection (FD) framework that collects packets from every flow at a pace proportional to the flow rate. To do so, the proposed FD scheme leverages low-frequency (i.e., three-way handshaking packets) and high-frequency (i.e., acknowledgment packets) phases of the transmission control protocol (TCP). This categorization enables speeding up the classification process and minimizes the associated delays. Furthermore, it has flexibility to be implemented either in a distributed (in-network) or centralized manner to classify flows of modifiable and unmodifiable hosts, respectively.

\item[$\bullet$] 
The in-network scheme is a module installed in virtual-switches/hypervisors, where the network is relaxed by utilizing location privileges of virtual-switches/hypervisors. On the other hand, the centralized scheme is implemented in the central unit (CU) by preconfiguring edge switches (ESs) to capture indicative packets (e.g., ACK, SYN, FIN, RST) which contain general information about a flow such as its starting time, source, and destination. The evaluation results show that the proposed centralized FD mechanism is about 224$\times$ faster than the existing centralized solutions with a 100\% accuracy with very low computational and communication overhead. Nonetheless, capturing and forwarding all the packets naturally introduce extensive processing and communication overhead. As a remedy, the ESs are configured to capture only indicative TCP packets. Two techniques are developed to further reduce the overhead: \textit{Stop Useless Notifications} and \textit{Sample Only ACK Packets}, where we tackle the challenge of differentiating between ACK and non-ACK packets. Noting that overhead of the centralized scheme without proposed techniques are close to that of the well-known sampled flow (sFlow) standard \cite{sflow}, developed methods provide up to 62$\times$ lower overhead.

\end{itemize}

\subsection{Paper Organization}
The rest of the paper is organized as follows: The background and related work is presented in Section \ref{sec:relatedwork}. The node architecture, network topology, and an overview of LightFDG is introduced in Section~\ref{sec:LightFDG}. Flow grooming and forwarding policy is explained in Section \ref{sec:FG}. Then, centralized and in-network FD schemes are provided in Section~\ref{sec::algorithm}. The Section~\ref{sec::overhead} presents the overhead minimization methods for the centralized scheme. Thereafter, the performance evaluations are presented in Section~\ref{sec:evaluation}. Finally, conclusions are drawn in Section~\ref{sec:conclusion}.

\section{Background and Related Work}
\label{sec:relatedwork}

\subsection{Literature Survey}
Even though traffic grooming has been extensively studied for passive optical networks \cite{Huang2007dynamic}, it was first introduced for wired DCNs in \cite{Sankaran2014scheduling, Sankaran2016optical} where flows are simply groomed into three classes of wavelengths. The groomed wavelengths are broadcasted within racks and higher-layer switches. However, these works only consider fixed wavelength capacity and do not deal with the flow characteristics of real-life DCNs. In \cite{Celik2018Design, Celik2019Design}, the authors conceptualized FG for MFs in WDM-FSO based wireless DCNs consisting of optoelectronic switches. After formulating the optimal FG problem, a suboptimal FG policy is designed for MFs, whereas EFs are carried out separately via server-to-server express lightpaths without going through any grooming operation. The results showed that the proposed FG method provides superior performance in both throughput and flow completion times. 

Existing FD schemes can be categorized into in-host~\cite{Mahout}, in-network~\cite{heavyhitter, sdc, opensketch} and centralized~\cite{hedera,  opensample, plank, everflow}. The in-host scheme is implemented in the host kernel where packets are counted before being injected into the network. When the counter exceeds a pre-defined threshold value, the flow is considered as an EF. However, in-host solutions need to modify the kernel network stack of the DCN hosts and hence require higher privileges that cannot always be granted. 

Unlike the in-host approaches, in-network schemes detect EFs at network switches. The flow information along with their counters are maintained at the switch memory whereas the comparison task is performed internally at the switch hardware (e.g., HashPipe~\cite{heavyhitter} and OpenSketch~\cite{opensketch}) or software (e.g., Software-Defined Counter~\cite{sdc}), or otherwise is exposed to a centralized entity (e.g., FlowRadar~\cite{flowradar}). One of the main challenges of the in-network mechanism is the limited resources of network switches (e.g., memory, processing delay, and power consumption). For instance, the time budget of DCN switches is too small (e.g., 25 ns) for a 128 bytes packet in a 40Gbps switch. Moreover, flow arrival-rates are in the order of milliseconds~\cite{flowchar} and the number of flow-entries in switching/routing tables is in the order of a hundred thousand. As a remedy, the researchers introduced hybrid solutions where some of the classification functions are migrated into cheaper hardware~\cite{opensketch} or into a centralized entity~\cite{flowradar}. 

The centralized schemes employ the available network features (e.g., packet sampling, port mirroring, or statistical polling) to collect network statistics and forward them to a centralized collector, which is usually programmed to perform the flow classification functions. However, the existing centralized schemes have limitations such as high monitoring overhead, the necessity for hardware modifications, high latency, and low accuracy~\cite{opensample, plank, everflow, sflow}. For example, OpenSample measures the link utilization from the sequence numbers in the TCP header and uses sFlow to collect the TCP packets \cite{opensample}. Planck \cite{plank} utilizes the port-mirroring, which causes extra processing overhead and cabling from  ESs to the centralized collector. Moreover, it is possible to miss some EFs due to the limited buffer size. Similar to sFlow, Planck can also attempt to detect already classified flows. To overcome these limitations, a hierarchical statistics pulling (HSP) mechanism is developed by utilizing a combination of aggregate and individual statistical messages in the OpenFlow protocol \cite{Lin2014elephant}. In FLight~\cite{flight}, centralized classification idea  is introduced with preliminary results extracted from small-scale experiments. Unfortunately, FLight fell short of convincing the idea can overcome the practical limitations of Planck.

\subsection{Distinctive Features of LightFDG}
LightFDG can be distinguished from above works in the following aspects:
\begin{itemize}
\item Instead of assuming flow classes are known a priori, LightFDG provides a holistic approach by integrating FGF with effective FD mechanisms. Numerical results show that classification accuracy and speed has a significant impact on the FGF performance. 

\item Unlike the above centralized and in-network FD schemes, LightFDG also provides a generic flow classification mechanism that is capable of operating either in centralized or in-network fashion. In addition to being fast-accurate-and-lightweight, the proposed detection scheme is favorable for two reasons: 1) To strike a balance between communication and computation overhead, and 2) To be able to compatible with both modifiable and unmodifiable hosts.  
\end{itemize}

\section{An Overview of The LightFDG}
\label{sec:LightFDG}
This section first explains the node architecture and DCN topology, then outlines the proposed LightFDG approach, and finally briefly introduce its algorithmic implementation.

\subsection{Node Architecture and Network Topology}
\begin{figure}[t!]
\begin{center}
 \includegraphics[width=1 \columnwidth]{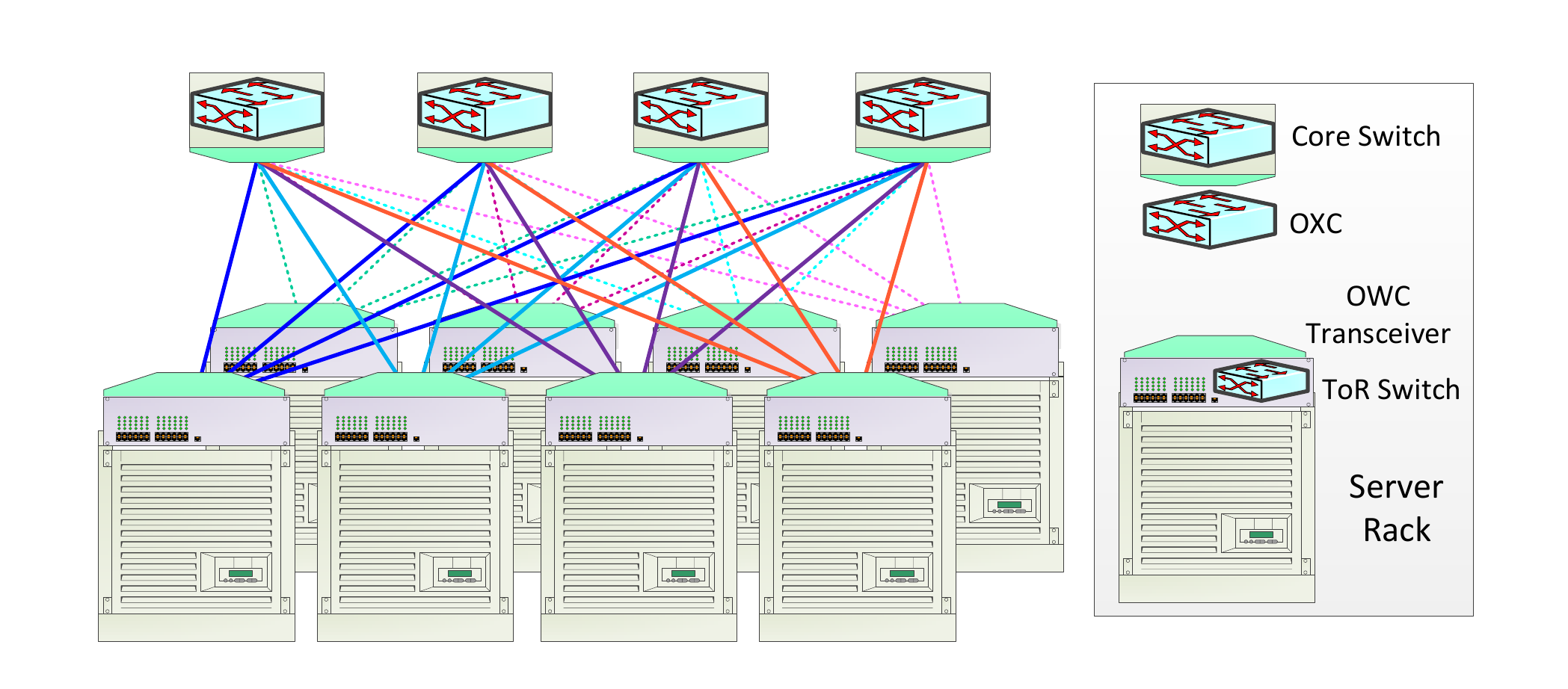}
\caption{Proposed topology for $N=8$ and spine/leaf ratio of $\eta=1/2$.}
\label{fig:top}
\end{center}
\end{figure}

A two-tier DCN architecture has been considered where each of $N$ leaf layer ESs is connected with $\eta N$ core switches (CSs) in the spine layer, where $\eta$ is the spine/leaf ratio. That is, leaf and spine layers are interconnected in a full-mesh manner as shown in Fig. \ref{fig:top}. LoS connectivity between these layers can be realized by using the physical topology in Fig. \ref{fig:top} where optical transceivers of ESs are connected to CS transceivers located at the top. WDM-FSO links can be implemented by photonic integrated circuits (PIC) \cite{photonics5030021} or photo-diode array \cite{Zhang2018} based transmitters to steer the lightbeam to the photo-diode receivers. Since transceivers are not required to change their direction after the setup, an alternative approach is setting gimbals atop the racks and directing transmitter and receiver towards each other\footnote{We should note that our contributions do not involve in hardware topology of WDCNs. We are rather interested in virtual topology design to investigate the flexibility and reconfigurability advantages of wireless DCNs.}. 

Each ES is considered as an optical cross-connect (OXC) with $N$ I/O ports such that received optical beam at each input port is first demultiplexed into $W$ wavelengths then fed into the connection matrix which determines the connectivity between OXC's I/O ports. After that output lines of the connection matrix are groomed via a multiplexer at each wavelength based on the proposed FG policy. Finally, the output of the OXCs are forwarded to the relevant optical transmitters as per the routing protocol. Since CSs are not involved in any grooming operation, they are simply modeled as optical routers to forward the R2R traffic. 

Thanks to the WDM, each link can be treated as a set of $W$ parallel FSO channels each of which is assumed to operate on intensity-modulation direct-detection (IM-DD) scheme. The channel capacity of wavelength $\omega \in [1,W]$ between optical transceivers $k$ and $l$ is given by \cite{Lapidoth2009capacity}
\begin{equation}
\label{eq:cap}
C_{kl}^\omega= B\frac{1}{2} \log \left(1+\frac{e (h_k^l)^2 (E_{k,l}^\omega)^2}{2 \pi} \right), \: \omega \in [1,W]
\end{equation} 
where $B$ is the bandwidth of a single wavelength, $E_{k,l}^\omega$ is the light intensity allocated to wavelength $\omega$, $h_k^l \triangleq \rho h_{k,l}^\ell h_{k,l}^a h_{k,l}^p$ is the optical channel gain which is assumed to be constant throughout a transmission block since the optical channel variations are very slow compared to the symbol duration \cite{Chaaban2017fundamental}, $\rho$ is the detector response, $h_{k,l}^\ell$ is the optical path loss, $h_{k,l}^a$ is the atmospheric turbulence, and $h_{k,l}^p$ is the pointing error. Due to hardware and safety concerns, signal intensity has to satisfy a total  intensity restrictions given by $\sum_{\omega} \mathrm{E}_{k,l}^\omega \leq \mathrm{E}_T$ , respectively.

\subsection{Integrated Flow Detection, Grooming, and Forwarding}
\begin{figure}[t!]
\begin{center}
 \includegraphics[width=1 \columnwidth]{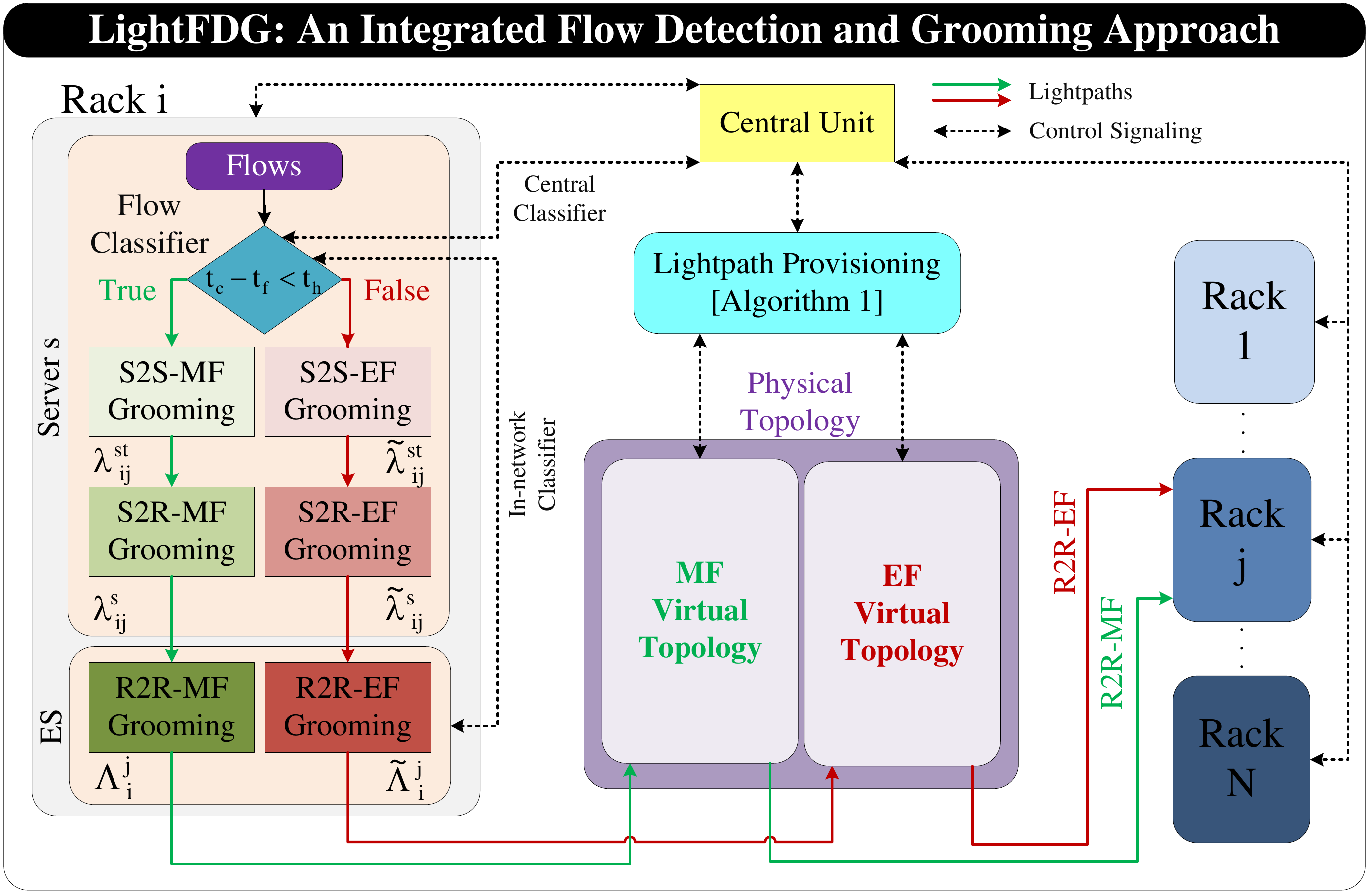}
\caption{Flowchart of the integrated flow detection and grooming approach.}
\label{fig:lightfdg}
\end{center}
\end{figure}

Before delving into the technical details, it is necessary to outline the principles of the proposed approach to integrate two main components: a) Flow detection (FD) and b) Flow grooming and forwarding (FGF).  For flows generated in the servers of rack $i$ and destined to the servers in rack $j$, LightFDG is illustrated in Fig. \ref{fig:lightfdg}  which is explained as follows:

\paragraph{Flow Detection} 
Since LightFDG is designed to handle different flow classes over non-overlapping virtual topologies separately, its first reaction to a new flow arrival must be identifying its class. The main motivation behind this disjoint processing lies within the differences of flow classes in terms of traffic characteristics (e.g., arrival rate, size, etc.) and quality of service (QoS) demands (e.g., flow completion time requests, delay sensitivity, bandwidth requirements, etc.). To handle distinct characteristics and meet the QoS demands of different classes, available network resources must be efficiently allocated to dedicated non-overlapping virtual topologies. 
 
As it is hardly possible to have a priori  flow class information upon the arrival, flows must be classified as soon as possible; otherwise, forwarding bandwidth hungry EFs over a path tailored to the needs of delay sensitive MFs may cause severe performance degradation \cite{hermes}. Therefore, to reap the full benefits of the flexibility and high fan-out attributes of WDM-FSO links, integrating FGF with fast-accurate-and-lightweight flow detection mechanisms is essential for QoS provisioning and efficient link utilization. As depicted in Fig. \ref{fig:lightfdg}, flow classifier of the LightFDG can be designed to operate either in distributed (i.e., in-network) or centralized fashion where classification is handled by ESs and CU, respectively. Regardless of the classifier type, LightFDG initially treats all incoming flows to be MF until the flow size exceeds a certain threshold. This is also intuitive since the DCN traffic characteristics tell us that the majority of the arriving flows are MFs whereas a significant portion of data is carried out by EFs. Once a flow is detected to be an EF, the source server immediately stops feeding its packets into the MF virtual topology and continue with the EF virtual topology instead\footnote{We must note that rerouting the flows to another path can cause re-sequence delay \cite{hermes}. Fortunately, negative impacts of this delay fades with the lifetime of EFs.}. 
 
\paragraph{Flow Grooming and Forwarding} 
Instead of handling each flow individually, grooming and forwarding aggregated traffic is a practical solution to avoid unnecessary topology reconfiguration delays and to eliminate high computational and control overhead. Hence, LightFDG separately executes flow grooming (FG) for each flow class as illustrated in Fig. \ref{fig:lightfdg} where flows are groomed in three steps to obtain R2R flows. These aggregated flows are then forwarded to the corresponding virtual topologies which are governed by the central unit (CU) based on a flow forwarding policy.  

Flow forwarding (FF) is responsible for forwarding R2R-MFs and R2R-EFs to the destination racks over predetermined R2R lightpaths of the corresponding virtual topology. A lightpath is defined as a pair of routing path and wavelength and is required to comply with the following constraints: 1) \textit{Collision constraint} dictates that a wavelength on a certain link cannot be shared by different lightpaths and 2) \textit{Wavelength continuity constraint} requires lightpaths to operate on the same wavelength along the routing path since OXCs are assumed to be not capable of wavelength conversion. Based on the lightpath provisioning procedure of Algorithm \ref{alg:LightFDG}, the CU manipulates virtual topologies to provision R2R lightpaths, which consists of three tasks: light intensity (i.e., capacity) allocation, wavelength assignment, and routing. 

\begin{algorithm}[t!]
\footnotesize
 \caption{\textbf{- LightFDG}}
  \label{alg:LightFDG}
\begin{algorithmic}[1]
 \renewcommand{\algorithmicrequire}{\textbf{Input:}}
 \renewcommand{\algorithmicensure}{\textbf{Output:}}
\Require $\mathcal{F} \left( \mathcal{V},\mathcal{W}, \mathcal{E} \right)$, $th$
 \State $\mathcal{F} \left( \mathcal{V},\mathcal{W}, \mathcal{E} \right) \gets $ Initialize the physical topology
 \State $\mathcal{G}_m \left( \mathcal{V}_m,\mathcal{W}_m, \mathcal{E}_m \right) \gets $ Initialize the MF virtual topology
  \State  $\mathcal{G}_e (\mathcal{V}_e, \mathcal{W}_e, \mathcal{E}_e)  \gets $ Initialize the EF virtual topology
 \State  $(\mathcal{L}_i^j , \mathcal{G}_m \left( \mathcal{V}_m,\mathcal{W}_m, \mathcal{E}_m\right)) \gets \textsc{Lightpath Provisioning}\:(m)$
  \State  $(\tilde{\mathcal{L}}_i^j, \mathcal{G}_e (\mathcal{V}_m, \mathcal{W}_m, \mathcal{E}_m)) \gets \textsc{Lightpath Provisioning}\:(e)$
\For{each flow $f$ between racks $i$ and $j$, $\forall i,j$,}
\State $t_f \gets$ Record the initial sequence number of flow $f$.
\While {$f$ is not completed}
\State $t_c \gets$ Update the sequence number
\If {$t_c-t_f < th$} (\textit{Flow Detection})
\State \textit{Flow Grooming:} Perform three-step MF grooming
\State \textit{Flow Forwarding:} Forward the R2R-MF over $\mathcal{L}_i^j$
\Else
\State \textit{Flow Grooming:}  Perform three-step EF grooming
\State  \textit{Flow Forwarding:}  Forward the R2R-EF over $\tilde{\mathcal{L}}_i^j$
\EndIf
\EndWhile
\EndFor
\vspace{2pt}
\hrule 
\vspace{2pt}
\Procedure{Lighpath Provisioning }{$x$}
\For{i=1:N}
\For{j=1:N}
\State $E \gets$ Determine the required intensity as per (2)%(\ref{eq:Ekl}).
 \State $\vect{\mathcal{P}}, \vect{\mathcal{C}} \gets$ Obtain K shortest paths and costs using K-SPR
\State $\mathcal{P}_{ij}^x \gets$ Decide on the path Rack$_i \rightarrow$Rack$_j$.
\State $\mathcal{W}_{ij}^x \gets$ Assign the wavelength to $\mathcal{P}_{i}^{j}$ 
\State $\mathcal{L}_{ij}^x  \gets ( E_{ij}^x , \mathcal{P}_{ij}^x  , \mathcal{W}_{ij}^x  )$ Record the lightpath Rack$_i \rightarrow$Rack$_j$.
 \State $\mathcal{G}_x \left( \mathcal{V}_x,\mathcal{W}_x, \mathcal{E}_x \right) \gets $ Update the virtual topology.
  \State $\mathcal{F} \left( \mathcal{V},\mathcal{W}, \mathcal{E} \right) \gets $ Update the physical topology.
 \EndFor
 \EndFor

\hspace{-9pt} \Return $\mathcal{G} \left( \mathcal{V},\mathcal{W}, \mathcal{E} \right)$, $\mathcal{L}_{ij}^x $, $\forall i, j$.
\EndProcedure
 \end{algorithmic}
 \end{algorithm}
\subsection{Algorithmic Implementation of LightFDG}
The pseudo-code of the LightFDG is given in Algorithm \ref{alg:LightFDG} where the first line initializes the virtual topology which is always kept updated by the CU over a wavelength dedicated for broadcasting the control signals. The MF (EF) virtual topology is characterized by a weighted graph $\mathcal{G}_x (\mathcal{V}_x,\mathcal{W}_x,\mathcal{E}_x)$ where $x=m$ for MFs, $x=e$ for EFs, $\mathcal{V}_x$ is the set of nodes, $\mathcal{E}_x$ presents available light intensity, and $\mathcal{W}_x$ symbolizes wavelength availability of the FSO links. Similarly, physical topology is denoted by $\mathcal{F} (\mathcal{V},\mathcal{W},\mathcal{E}) \supseteq \bigcup_x \mathcal{G}_x (\mathcal{V}_x,\mathcal{W}_x,\mathcal{E}_x)$. After that, the algorithm first provisions the R2R-MF lightpaths between the rack pairs (line 2). The residual light intensity and wavelengths are then exploited to provision the R2R-EF lightpaths between the racks. Since LightFDG guarantees the QoS demands of delay-sensitive MFs first, they are prioritized over EFs for protection against performance degradation in heavy traffic conditions. The details of lightpath provisioning are provided in the next sections. 

For each flow arrival, an initial ACK sequence number, $t_f$, is recorded along with the current sequence number, $t_c$ (lines 6-8). As long as the packet size is below a certain threshold, $t_c-t_f<th$, LightFDG regards flow $f$ as an MF (line 10), grooms it with other MFs destined to the same rack (line 11), and forwarded over predetermined R2R-MF lightpaths (line 12). Otherwise, $f$ is regarded as an EF, groomed it with other EFs destined to the same rack (line 14), and forwarded over predetermined R2R-EF lightpaths (line 15). In the remainder, we present more technical details regarding each step of Algorithm \ref{alg:LightFDG}.

\section{Flow Grooming and Forwarding Policy}
\label{sec:FG}
FGF policy comprises of three joint subproblems: 1)  Designing grooming strategy,  2) Virtual topology design, i.e., provisioning lightpaths' capacity and routing the aggregated traffic over the physical topology, and 3) Assignment of wavelengths to the lightpaths. Since each of these subproblems are NP-hard \cite{Zhur2002mesh}, FGF is also an NP-hard problem which belongs to the class of mixed-integer linear programming (MILP) problems. Due to dynamic and heavy traffic characteristics of DCNs, obtaining an optimal FGF policy requires impractical time complexity even for small-scale DCNs. Hence, fast yet high-performance suboptimal solutions are necessary to implement FGF for WDCNs.

\subsection{Flow Grooming: A Three-Step Strategy}
The three-step optical grooming policy aggregates each class of flows in the optical domain as follows  \cite{Celik2018Design,Celik2019Design} :
\begin{enumerate}
\item  \textit{Server-to-server (S2S) Grooming:} Each server combines all flow arrivals destined to the same server into a single flow. Denoting the set of hosts/servers within rack $i$ as $\mathcal{R}_i$, arrival rates of S2S-MFs and S2S-EFs from  $s \in \mathcal{R}_i$ to  $t \in \mathcal{R}_j$ are assumed to follow Poisson distribution with arrival rates $\lambda_{ij}^{st}$ and $\tilde{\lambda}_{ij}^{st}$, respectively. 

\item \textit{Server-to-Rack (S2R) Grooming:} Servers further groom S2S flows according to common destination rack such that all flows destined to the same rack is groomed into a single S2R flow. Assuming that flow arrivals are independent from each other, S2R-MF and SR-EF arrival rates also follow a Poisson distribution with the composite rates of $\lambda_{ij}^s=\sum_{t\in \mathcal{R}_j}  \lambda_{ij}^{st}$ and $\tilde{\lambda}_{ij}^s=\sum_{t\in \mathcal{R}_j}  \tilde{\lambda}_{ij}^{st}$, respectively. 

\item  \textit{Rack-to-Rack (R2R) Grooming:} The ES finally grooms all S2R flows received from different servers according to their destination racks to obtain R2R flows. Similarly, overall arrival rate for the R2R-MF and  R2R-EF flows have the composite rates of $\Lambda_{ij}^m=\sum_{s\in \mathcal{R}_i}\sum_{t\in \mathcal{R}_j} \lambda_{ij}^{st}$ and $\Lambda_{ij}^e=\sum_{s\in \mathcal{R}_i}\sum_{t\in \mathcal{R}_j} \tilde{\lambda}_{ij}^{st}$, respectively. 
\end{enumerate}
Following the three-step grooming, R2R-MF and R2R-EF flows are then directed to the relevant optical transmitter based on the predetermined routing paths. Therefore, CSs are not required to implement any optical grooming since they merely forward the lightpaths to the ES transceivers of the destination racks. 

\subsection{Flow Forwarding: Lightpath Provisioning Procedure}
We denote the R2R lightpaths between racks $i$ and $j$ by $\mathcal{L}_{ij}^x=\{E_{ij}^x, \mathcal{P}_{ij}^x, \mathcal{W}_{ij}^x\}$ where $x=m$ ($x=e$) represents MFs (EFs), ${E}_{ij}^x$ is the allocated light intensity, $\mathcal{P}_{ij}^x$ is the routing path, and $ \mathcal{W}_{ij}^x$ is the assigned wavelength. As shown between lines 19 and 28, we iteratively provision lightpaths for the rack pairs as follows. 

The first step of the lightpath provisioning (line 22) is determining the required light intensity of the wavelengths along the R2R paths, which is explained as follows: Based on the flow completion time $\tau_m$ and flow size $F_m$ of MFs, required capacity for R2R-MF between Rack$_i$ and Rack$_j$ is given by, $C_{ij}^m=\Lambda_{ij}^m F_m/ \tau_m$, which must be guaranteed at each hop along the routing path. By substituting $C_i^j$ into \eqref{eq:cap}, required light intensity for wavelength and link pairs of $\mathcal{L}_{ij}^m $  is given by
\begin{equation}
\label{eq:Ekl}
E_{k,l}^\omega=\left(  \frac{ \left( 2^{\frac{2 C_{ij}^m }{B}}-1\right)  2 \pi }{ e(h_k^l)^2 } \right)^{1/2}, \forall k,l \in \mathcal{P}_{ij}^m , \omega \in \mathcal{W}_{ij}^m .
\end{equation} 
Likewise, required light intensity for wavelength and link pairs of $\mathcal{L}_{ij}^e$ can be obtained by substituting $C_{ij}^e =\Lambda_{ij}^e F_e/\tau_e$ into \eqref{eq:cap}.

In Line 23, we eliminate edges with insufficient light intensity (i.e., availability is less than the required amount in \eqref{eq:Ekl}) and run a modified version of the loop-less k-shortest path (KSP) algorithm \cite{KSP} which not only finds the path with the maximum available light intensity (i.e., capacity), but also $k-1$ other paths in non-increasing order of the available intensity. Thereafter, the available light intensity of these $k$ lightpaths are weighted with the available number of wavelengths on the route. Finally, line 24 selects the lightpath with the maximum intensity and number of wavelength product as the dedicated R2R path for the racks $i$ and $j$. In this way, we ensure at each iteration that the number of available wavelengths and light intensity on FSO links are evenly distributed across the entire topology. 

In order to ensure that there exists no less than two R2R paths between each rack pairs (i.e., one for MFs and other for EFs), the required least number of wavelengths is given by  
\begin{equation}
\label{eq:W}
W \geq \left \lceil \frac{2(N-1)}{\eta N}  \right \rceil.
\end{equation}
Assuming \eqref{eq:W} holds, line 25 randomly assigns wavelengths to lightpaths since they are identical once the lightpath capacity and route is determined. Nonetheless, we must note that wavelength continuity and collision constraints may cause connectivity issues between the rack pairs if the number of wavelengths is limited (i.e., $W < \left \lceil 2(N-1)/ (\eta N)  \right \rceil)$ \cite{TRUONGHUU2016184}. In this case, lightpath provisioning procedure must allow multiple lightpaths to share the same wavelength based on a certain multiple access scheme, e.g., time-division multiple access \cite{Dixit2012TDMAWDM}. After that, line 26 records the lightpath and line 27 (28) updates the virtual (physical) topology for the next iteration. Following the MF lightpath provisioning in line 3, line 4 also employs the above \textsc{ lightpath provisioning} procedure for R2R-EF lightpaths by exploiting the residual light intensity and available wavelengths on FSO links.

 \section{Flow Detection: A Fast, Accurate, and Lightweight Framework}
\label{sec::algorithm}
As the TCP carries out almost 99\% of the DCN traffic \cite{flowchar}, its attributes are leveraged by many existing flow classification methods to build accurate and fast flow detection mechanisms \cite{Mahout, opensample}. A single packet of every TCP flow offers partial information about the flow size which is useless if it is individually considered. When two packets are captured from the same TCP flow, one can discover the number of bytes that have been transmitted between the two captured packets. This technique is fundamental idea of the existing flow detection solutions (e.g., OpenSample-TCP and Planck) where the difference between the TCP sequence numbers of captured packets and time of capture is used to measure the link utilization. 

\subsection{Operational Principles of the Proposed FD Schemes}
LightFDG divides the TCP communication into low-frequency and high-frequency phases. The former consists of three-way handshaking packets (i.e., SYN, SYN+ACK, and ACK) that appears once to set the initial sequence number and prepare the connection parameters at the beginning of each flow. On the other hand, the latter comprises of ACK packets sent by the receiver to acknowledge the transmitter about the successful reception of the flow packets. LightFDG also benefits from the privileges of its position, i.e., the CU in the centralized scheme or the virtual-switch/hypervisor in the in-network scheme.

To this end, LightFDG employs two main components: collector and classifier. The collector reads the initial sequence number (ISN) of every TCP flow, which is the first half of the required information to classify a flow, and stores it in the flow-information table. Likewise, the classifier needs to compare the ISN with every captured packet from the same flow until a threshold value is reached. However, instead of reading/capturing\footnote{We use the term \textit{reading} for in-network scheme and \textit{capturing} for the centralized scheme.} the data packets themselves, LightFDG is programmed to read/capture the headers of ACK packets. For instance, if the classifier reads ACK$_{1}$ and then ACK$_{t}$ after a while, the number of transmitted bytes between these two ACKs is calculated as $t - 1$. The ACK sequence number only presents the bytes that have been successfully received, that is, the lost or out-of-order packets are not counted. Therefore, the classifier needs only to compare the number of bytes with a predefined threshold value $th$ to classify a flow as an EF. Once $t-1>th$ is satisfied, the flow is considered as an EF and the subsequent packets are marked to be forwarded via the virtual topology of EFs and the CU is reported about this detection.

Even though we consider one-way flows throughout the paper, the LightFDG is also capable of handling two-way flows where the receiver temporarily delays sending ACKs until it has data to be sent towards the transmitter. This is also known as piggybacking and favored by its efficient bandwidth utilization \cite{tanenbaum2011computer}. In principle, one-way LightFDG can still operate under piggybacking communication by individually handling flows at both directions. That is, the CU updates two records in the flow information table, i.e., one for each direction. Whenever a direction exceeds the predefined threshold value, its packets are marked to be forwarded through the EF virtual topology. However, delaying ACKs may cause the receiver jamming the service if it has no data to send back. This delay will naturally reduce the flow detection speed of LightFD. Fortunately, this issue can be resolved by enabling a receiver timeout counter when a data frame is received. If the count terminates and there is no data frame to send, the receiver will send an ACK. The sender also adds a emitter timeout counter, if the counter ends without receiving confirmation, the sender assumes packet loss, and sends the frame again \cite{tanenbaum2011computer}.

\subsection{In-Network and Centralized FD Schemes} 
Although centralized and in-network schemes use the same components (i.e., collector and classifier), they differ in the implementation such that the centralized scheme is implemented in a CU (e.g.,  SDN controller) whereas the in-network scheme is implemented as a kernel module in the hypervisor/virtual-switches. While the centralized scheme is used to classify the flows of unmodifiable hosts, the in-network scheme is used to classify the flows of modifiable hosts. Due to its position in the hypervisor/virtual-switch, the in-network is faster than the centralized scheme in terms of detection speed. Although the administrative overhead of the in-network scheme is lighter than that of the in-host scheme, it needs to be installed in every hypervisor/virtual-switch, and the DC is expected to have various versions and brands. On the other hand, the centralized scheme is simpler where the network administrator needs only to install it in the CU, and it will configure the switches with the needed configurations. In what follows, we provide the implementation details of LightFDG in both centralized and in-network cases.

  \begin{figure}[t]
  \centering
  \includegraphics[width=0.8\linewidth]{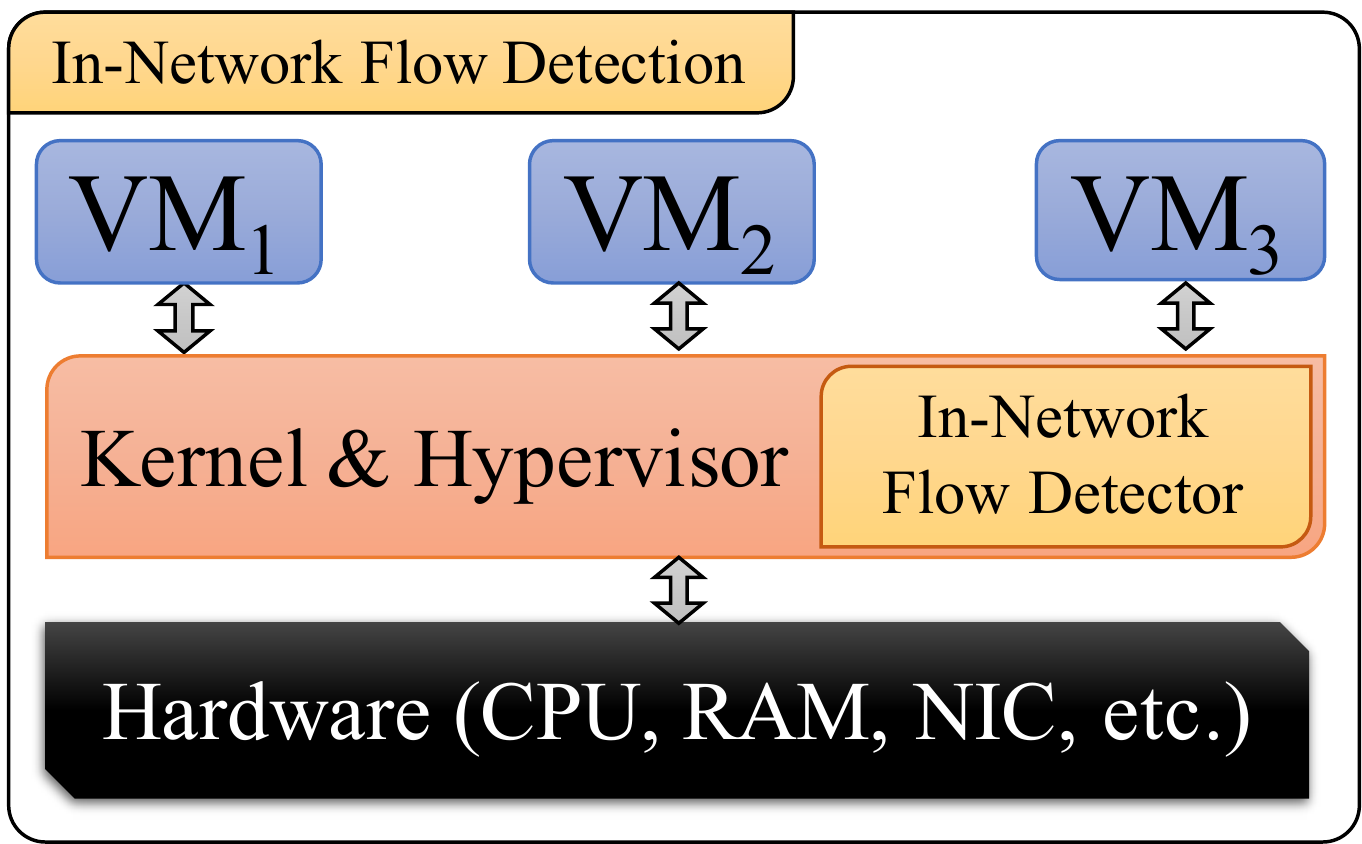}
 \caption{The in-network FD scheme installed in the hypervisor.}
  \label{fig::in-network}
\end{figure}
 
\subsubsection{In-network Scheme}\label{sec::in-network}
As shown in Fig.~\ref{fig::in-network}, the in-network LightFDG scheme is designed as module installed on the kernels of hypervisor of edge servers or in the virtual switches. The kernels of these systems provide a degree of control on exchanging packets, i.e., access to their packet headers and unencrypted payload. For EF detection, LightFDG exploits two main components: collector and classifier. The collector is responsible for learning general flow information from the messages of the low-frequency phase. Moreover, it has a flow-information table to store the flow information (e.g., source/destination IP/MAC addresses and TCP sequence numbers). On the other hand, the classifier is accountable for detecting EFs from headers of the high-frequency phase messages.

 \begin{figure}[t]
  \centering
    \includegraphics[width=\linewidth]{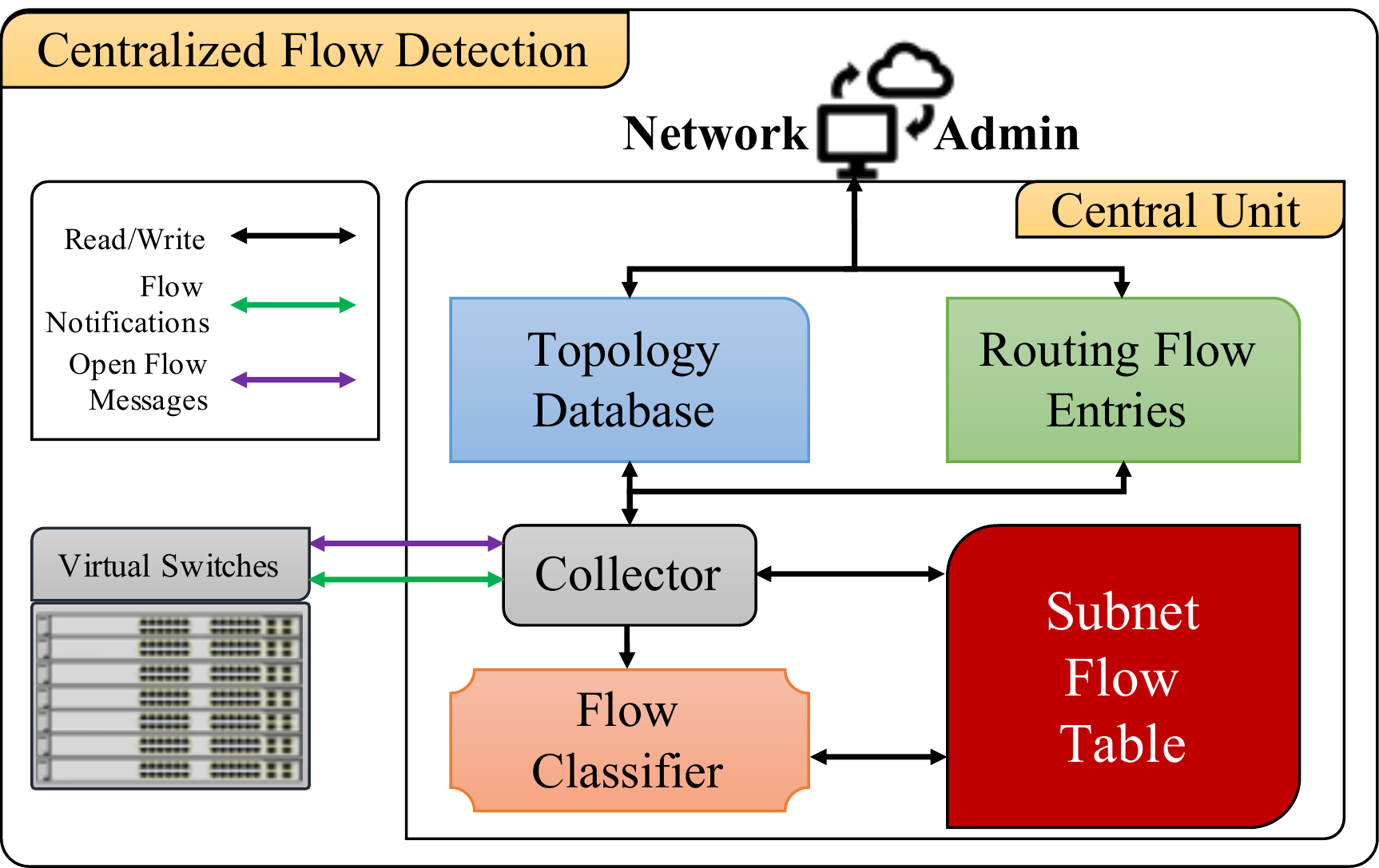}
 \caption{Flowchart of the centralized FD scheme.}
  \label{fig::design}
\end{figure} 
 \subsubsection{Centralized Scheme}\label{sec::central}
 
The network switch/router configurations can be performed physically or remotely: In the former, portable machines are connected to console port on the switch/router to perform the configuration steps, which is repeated for every switch/router in the network. In the latter, configurations are remotely written by using telnet/ssh terminal over physical links established between every switch/router and the CU after the necessary remote configuration activation commands are physically entered into the switches/routers. Alternatively, LightFDG employs a programmable interface (i.e., OpenFlow) that remotely communicates with ESs and configure them to capture the low and high frequency phase packets. Unlike the aforementioned methods, OpenFlow protocol has more abstracted and advanced commands that facilitate the configuration of network devices. Taking the dynamic nature of FGF into account, OpenFlow protocol is also a key solution to create virtual topologies and reconfigure it based on changes driven by FGF policy.

As demonstrated in Fig.~\ref{fig::design}, the centralized FD scheme configures ES to capture the TCP first phase packets (i.e., SYN+ACK and RST/FIN) and ACKs within the second phase packets. When these packets are present in a flow, the ESs send a \textit{flow-notification} to collector of the CU which reads and writes the flow information and ISNs in a flow-information table. To accelerate the classification process, the flow-information table has been divided into multiple small tables each of which is indexed by a subnet ID. When the collector receives a captured message, it needs to figure out which table this message belongs to. Since every OpenFlow header has the ID of the source switch and the flow information, it can use the source IP or the data-plane ID to achieve this task\footnote{In this work, we prefer to use the data-plane ID in the OpenFlow messages.}. To determine how many bytes have been successfully received, the flow classifier compares the sequence number of captured ACKs, $t_c$, with their correspondent ISNs in the subnet flow-information table. When the number of bytes exceeds the EF threshold value, the flow is then considered as an EF.

 \begin{figure}[t]
  \centering
    \includegraphics[width=0.8\linewidth]{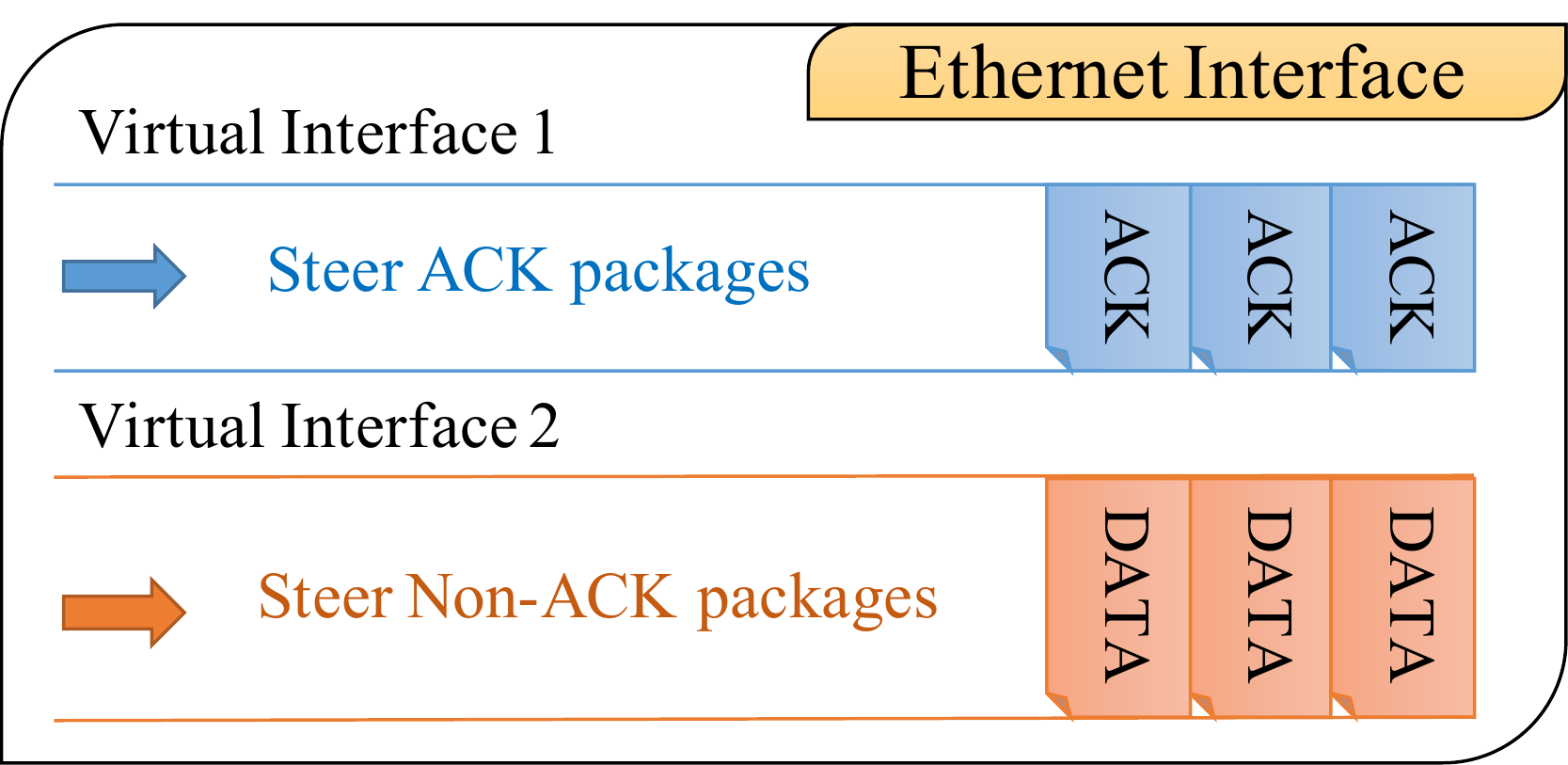}
 \caption{The LightFDG steers the ACK and non-ACK packets to separate virtual interfaces on every ES.}
  \label{virtualint}
\end{figure}

\section{Minimizing the Network Overhead} \label{sec::overhead}

The propagation of low and high frequency phase packets through the DCN may introduce a queuing delay and processing overhead on the network. Before proceeding to the proposed overhead reduction methods, it is important to demonstrate the feasibility of the fundamental idea by investigating two kinds of datasets; DCN \cite{flowchar} and enterprise~\cite{inter} datasets. In Table ~\ref{table:ack_size}, the collections are  ordered  such that the sequence number is a suffix added to the name of the datasets. Table ~\ref{table:ack_size} shows that normal LightFDG without overhead minimization techniques (i.e., capturing only the low and high frequency packets) delivers a low overhead in comparison with capturing all the packets. Although LightFDG is required to collect much less packets then the total number of packets in DCN datasets, enterprise dataset may cause a significant overhead since almost all packets are required to be captured at some cases. Accordingly, following overhead reduction methods are proposed:

\subsection{Sample Only ACKs Packets}
In practice, there are two methods to capture and notify the classifier for every ACK message: 1) Configuring the ESs with particular OpenFlow rules to mirror every ACK message to the CU and 2) Using sFlow to sample every packet. Although the former method has a high accuracy, mirroring every ACK message to the collector introduces high communication and processing overhead. On the other hand, the random packet sampling has a low overhead and at the same time low accuracy. Therefore, the number of sampled packets is similar to the number of sampling from every TCP packet. To get the best of both methods, their advantages are combined as shown in Fig.~\ref{virtualint} where sampling is only done for ACK packages by steering them to a separate virtual interface. In this way, the unnecessary overhead caused by sampling data (non-ACK) packets are eliminated. These interfaces could be a high priority queue or a virtual interface associated with every interface in the ESs. Here, we especially focus on the ESs to reduce the overhead of capturing multiple packets of the same flow during a short period.

\subsection{Stop Useless Notifications}
The useless notifications are the messages that lost their value for the flow classifier such as the ACK messages of the already classified flows or notifications regarding flows that can be classified from their packet headers.  Accordingly, the proposed FD mechanism relieves the overhead in two ways: 
\begin{enumerate}
\item
It is widely known that some portion ($>$30\%) of flows can be pre-classified from application types (e.g., FTP, Syslog, and NTP) \cite{trafficsurvey,trafficclass1,trafficclass2}. Indeed, LihtFDG is capable of pre-classification based on application types because it already captures low-frequency phase packets. At the beginning of each flow, LightFDG can immediately and accurately identify the flow class of these applications from TCP port numbers~\cite{trafficsurvey}. This naturally eliminates the need for capturing high-frequency phase packets and resulting overhead.

\item  Furthermore, the CU is triggered to reconfigure the ESs to stop capturing its ACK messages when the classifier detects an EF. In OpenFlow, for instance, the controller is responsible to perform such actions and install the necessary flow-entries with higher priority on the ESs. In this case, the classifier will not receive the notification messages about the classified flows, which naturally reduces the overhead.
\end{enumerate}

%%%%%%%%%%%%%%%%%%%%%%%%%%%%%%%%%%%%%%%%%%%%%%%%%%%%%%%Table
\begin{table}[t]
\scriptsize
\captionsetup{justification=centering}
\caption{\\The impact of LightFDG on real packet traces. } % title of Table
\centering % used for centering table
\resizebox{\columnwidth}{!}{
\begin{tabular}{c c c c c} % centered columns (4 columns)
\hline\hline \\[0.01ex]%inserts double horizontal lines
Dataset(ID) & \# Packets & \# Flows & Normal LightFDG & Ratio \\ [0.5ex]
\hline \\ [0.001ex]% inserts single horizontal line
Inter. 1 & 5700526 & 662 & 2313433 & =2.5$\times$\\[1ex]
Inter. 2 & 2261261 & 3512 & 2174151 & =1.04$\times$ \\[1ex]
Inter. 3 & 84574 & 172 & 82817 & =1.02$\times$ \\[1ex]
Univ. 2.0 & 11917501 & 589 & 6333 & =1881.8$\times$ \\ [1ex]
Univ. 2.1 & 11917527 & 775 & 14636 & =814.3$\times$ \\ [1ex]
Univ. 2.3 & 11915859 & 727 & 7787 & =1530$\times$ \\[1ex]
Univ. 2.4 & 11915613 & 844 & 10528 & =1131.8$\times$ \\[1ex]
\hline %inserts single line
\end{tabular}
}
\label{table:ack_size} % is used to refer this table in the text
\vspace{-4 mm}
\end{table}

\begin{figure*}
    \centering
    \begin{subfigure}[b]{0.32\textwidth}
\definecolor{mycolor1}{rgb}{0.00000,0.50000,1.00000}%
\definecolor{mycolor2}{rgb}{1.00000,0.80000,0.00000}%
\begin{tikzpicture}

\begin{axis}[%
width=1.8in,
height=1.4in,
at={(0.994in,0.435in)},
scale only axis,
log origin=infty,
xmin=0,
xmax=5,
xtick={0,1,2,3,4,5},
xticklabels={\empty},
xticklabel style={rotate=45},
ymin=10,
ymax=110,
ylabel style={font=\tiny},
ylabel={Perecentage (\%)},
axis background/.style={fill=white},
xmajorgrids,
ymajorgrids,
legend style={at={(0.5,1.03)},line width=0.5pt, font=\tiny, anchor=south, legend columns=2, legend cell align=left, align=left, draw=white!15!black}
]
\addplot[ybar, bar width=0.3cm, fill=mycolor1, draw=black, area legend] table[row sep=crcr] {%
1	100\\
};
\addlegendentry{In-network}
\addplot[ybar, bar width=0.3cm, fill=mycolor2, draw=black, area legend] table[row sep=crcr] {%
2	100\\
};
\addlegendentry{Centralized}
\addplot[ybar, bar width=0.3cm, fill=lime, draw=black, area legend] table[row sep=crcr] {%
3	19\\
};
\addlegendentry{sFlow}
\addplot[ybar, bar width=0.3cm, fill=red, draw=black, area legend] table[row sep=crcr] {%
4	90\\
};
\addlegendentry{OpenSample-TCP}
\end{axis}
\end{tikzpicture}%
        \caption{100 Flows}
        \label{fig:100}
    \end{subfigure}
    \begin{subfigure}[b]{0.32\textwidth}
\definecolor{mycolor1}{rgb}{0.00000,0.50000,1.00000}%
\definecolor{mycolor2}{rgb}{1.00000,0.80000,0.00000}%
\begin{tikzpicture}

\begin{axis}[%
width=1.8in,
height=1.4in,
at={(0.994in,0.435in)},
scale only axis,
log origin=infty,
xmin=0,
xmax=5,
xtick={0,1,2,3,4,5},
xticklabels={\empty},
xticklabel style={rotate=45},
ymin=10,
ymax=110,
ylabel style={font=\tiny},
ylabel={Perecentage (\%)},
axis background/.style={fill=white},
xmajorgrids,
ymajorgrids,
legend style={at={(0.5,1.03)},line width=0.5pt, font=\tiny, anchor=south, legend columns=2, legend cell align=left, align=left, draw=white!15!black}
]
\addplot[ybar,   bar width=0.3cm, fill=mycolor1, draw=black, area legend] table[row sep=crcr] {%
1	100\\
};
\addlegendentry{In-network}
\addplot[ybar,  bar width=0.3cm, fill=mycolor2, draw=black, area legend] table[row sep=crcr] {%
2	100\\
};
\addlegendentry{Centralized}

\addplot[ybar,   bar width=0.3cm, fill=lime, draw=black, area legend] table[row sep=crcr] {%
3	30\\
};
\addlegendentry{sFlow}

\addplot[ybar,   bar width=0.3cm, fill=red, draw=black, area legend] table[row sep=crcr] {%
4	94\\
};
\addlegendentry{OpenSample-TCP}

\end{axis}
\end{tikzpicture}%
        \caption{500 Flows}
        \label{fig:500}
    \end{subfigure}
    \begin{subfigure}[b]{0.32\textwidth}
\definecolor{mycolor1}{rgb}{0.00000,0.50000,1.00000}%
\definecolor{mycolor2}{rgb}{1.00000,0.80000,0.00000}%
\begin{tikzpicture}
\begin{axis}[%
width=1.8in,
height=1.4in,
at={(0.994in,0.435in)},
scale only axis,
log origin=infty,
xmin=0,
xmax=5,
xtick={0,1,2,3,4,5},
xticklabels={\empty},
xticklabel style={rotate=45},
ymin=10,
ymax=110,
ylabel style={font=\tiny},
ylabel={Perecentage (\%)},
axis background/.style={fill=white},
xmajorgrids,
ymajorgrids,
legend style={at={(0.5,1.03)},line width=0.5pt, font=\tiny, anchor=south, legend columns=2, legend cell align=left, align=left, draw=white!15!black}
]
\addplot[ybar, bar width=0.3cm, fill=mycolor1, draw=black, area legend] table[row sep=crcr] {%
1	100\\
};
\addlegendentry{In-network}
\addplot[ybar, bar width=0.3cm, fill=mycolor2, draw=black, area legend] table[row sep=crcr] {%
2	100\\
};
\addlegendentry{Centralized}

\addplot[ybar, bar width=0.3cm, fill=lime, draw=black, area legend] table[row sep=crcr] {%
3	30\\
};
\addlegendentry{sFlow}

\addplot[ybar, bar width=0.3cm, fill=red, draw=black, area legend] table[row sep=crcr] {%
4	94\\
};
\addlegendentry{OpenSample-TCP}

\end{axis}
\end{tikzpicture}%
        \caption{1000 Flows}
        \label{fig:1000}
    \end{subfigure}
    \caption{Accuracy comparison of FD schemes in different traffic configurations during the mix scenario.}\label{fig:accuracy}
\end{figure*}
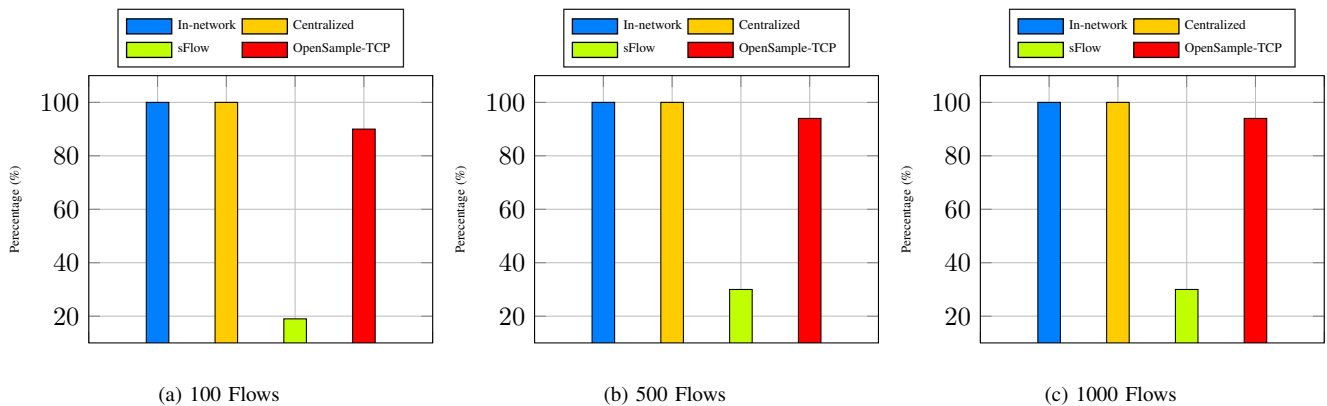

\section{Performance Evaluations}
\label{sec:evaluation}
In this section, we first evaluate the performance of the proposed FD mechanisms and show how they can improve the overall DCN throughput and reduce flow completion times (FCTs). To demonstrate the light overhead of LightFDG in a real DCN environment,  LightFDG is directly run on the same datasets with the minimization methods explained in Sec.~\ref{sec::overhead}. Thereby, the LightFDG performance is influenced by the network configurations (e.g., QoS settings) and the link loads of the datasets. The following subsections show the classification characteristics of LightFDG during different traffic loads and communication scenarios. Finally, the virtues of LightFDG in improving throughput and FCT performance are illustrated in comparison with other schemes.

\subsection{Network Setup}

Evaluations are conducted using Mininet emulator~\cite{mininet} and POX~\cite{pox} controller on machine with 16 $\times$ (2.5GHz-Intel Xeon CPU E5-2680v3) processors and 128 GiB memory. A leaf-spine topology is created with eight spines and leaves, each with 40 hosts. Iperf is used to generate flows whose arrivals follow exponential distribution. The Iperf permits specifying the flow size and communicating units. The hosts in Mininet are divided into two separate sets (clients and servers) and Iperf is configured to loop these two sets till all the members are visited and the launched flows are completed. The clients and the servers are always selected from different racks, and the number of MFs and EFs are determined according to the required percentage (e.g., 10\%, 15\%, and 20\%). The flow ID of flows are directly collected from the Mininet and stored in another file to compare them with the output results from the classifier. Our evaluation is based on network instances (one-shot), where a new instance is run after ensuring that all flows of the ongoing instance are complete. Linux monitoring features are used to monitor the processing of evaluation components such as Iperf PID of every running flow.

Although the CU and the network management servers of a real-life DC have high computational power to process large number of flows, such processing capacity can not be offered in a contained lab environment. To mimic the traffic load of real DCs, rather than randomly generating thousands of flows, different traffic loads are steered into a specific link where our classification and the competitor solutions are executed. In this respect, the traffic between the first and second subnet is steered on a single path to make sure that the number of flows on that path is as needed, i.e., 100, 500, and 1K. During the flow mix evaluations, the percentage of MFs is set to 90\% of the total flows. Unless stated explicitly otherwise, the size of MFs and EFs are 100KB and 128MB, respectively. EF detection threshold is 1MB.
For comparison purposes, we consider OpenSample-TCP and sFlow as our benchmark detection schemes and their sampling-rate is configured to 1-to-1000 as suggested in~\cite{DevoFlow, plank, opensample}. For the sampling rate of the ``Sample Only ACK Packets'' method proposed in Section VI-A, there is no clear and direct way of deciding on the sampling rate. Thus, different sampling rates are evaluated from 1-to-50 to 1-to-1000, and 1-to-100 is selected as it gives a desirable performance in terms of accuracy, speed and overhead. For the sake of a fair comparison with the other detection schemes, presented results do not consider any pre-classifiable flows to evaluate the performance of LightFDG at the worst-case. Nonetheless, it is worth to point out that overall LightFDG performance may be further improved with the increasing portion of pre-classifiable flows. At the extreme case of having all flows are pre-classifiable, the LightFDG will not suffer from classification delay and accuracy.

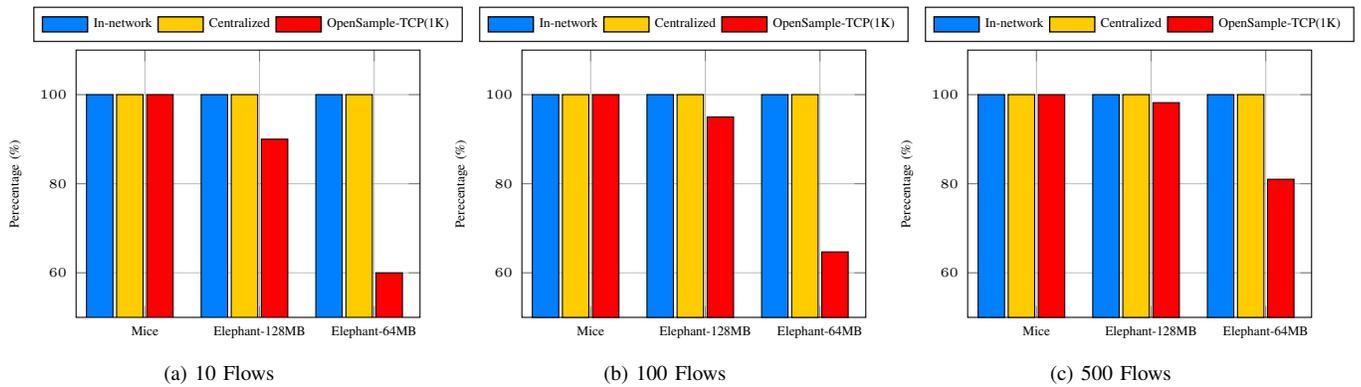
\begin{figure*}
    \centering
    \begin{subfigure}[b]{0.32\textwidth}
\definecolor{mycolor1}{rgb}{0.00000,0.50000,1.00000}%
\definecolor{mycolor2}{rgb}{1.00000,0.80000,0.00000}%
\begin{tikzpicture}

\begin{axis}[%
width=1.8in,
height=1.4in,
scale only axis,
log origin=infty,
xmin=0.4,
xmax=3.4,
xtick={1,2,3},
xticklabels={{Mice},{Elephant-128MB},{Elephant-64MB}},
tick label style={font=\tiny},
ymin=50,
ymax=110,
ylabel style={font=\tiny},
ylabel={Perecentage (\%)},
axis background/.style={fill=white},
xmajorgrids,
ymajorgrids,
legend style={at={(0.5,1.03)},line width=0.5pt, font=\tiny, anchor=south, legend columns=3, legend cell align=left, align=left, draw=white!15!black}
]
\addplot[ybar, bar shift=-.6cm, fill=mycolor1, draw=black, area legend] table[row sep=crcr] {%
1	100\\
2	100\\
3   100\\
};
\addlegendentry{In-network}

\addplot[ybar, bar shift=-.2cm, fill=mycolor2, draw=black, area legend] table[row sep=crcr] {%
1	100\\
2	100\\
3	100\\
};
\addlegendentry{Centralized}

\addplot[ybar, bar shift=.2cm, fill=red, draw=black, area legend] table[row sep=crcr] {%
1	100\\
2	90\\
3	60\\
};
\addlegendentry{OpenSample-TCP(1K)}

\end{axis}
\end{tikzpicture}%
        \caption{10 Flows}
        \label{fig:me100}
    \end{subfigure}
    \begin{subfigure}[b]{0.32\textwidth}
\definecolor{mycolor1}{rgb}{0.00000,0.50000,1.00000}%
\definecolor{mycolor2}{rgb}{1.00000,0.80000,0.00000}%

\begin{tikzpicture}

\begin{axis}[%
width=1.8in,
height=1.4in,
at={(1.752in,0.688in)},
scale only axis,
log origin=infty,
xmin=0.4,
xmax=3.4,
xtick={1,2,3},
xticklabels={{Mice},{Elephant-128MB},{Elephant-64MB}},
tick label style={font=\tiny},
ymin=50,
ymax=110,
ylabel style={font=\tiny},
ylabel={Perecentage (\%)},
axis background/.style={fill=white},
xmajorgrids,
ymajorgrids,
legend style={at={(0.5,1.03)},line width=0.5pt, font=\tiny, anchor=south, legend columns=3, legend cell align=left, align=left, draw=white!15!black}
]
\addplot[ybar, bar shift=-.6cm, fill=mycolor1, draw=black, area legend] table[row sep=crcr] {%
1	100\\
2	100\\
3	100\\
};
\addlegendentry{In-network}

\addplot[ybar, bar shift=-.2cm, fill=mycolor2, draw=black, area legend] table[row sep=crcr] {%
1	100\\
2	100\\
3	100\\
};
\addlegendentry{Centralized}

\addplot[ybar, bar shift=.2cm, fill=red, draw=black, area legend] table[row sep=crcr] {%
1	100\\
2	95\\
3	64.7\\
};
\addlegendentry{OpenSample-TCP(1K)}

\end{axis}
\end{tikzpicture}%
        \caption{100 Flows}
        \label{fig:me500}
    \end{subfigure}
    \begin{subfigure}[b]{0.32\textwidth}
\definecolor{mycolor1}{rgb}{0.00000,0.50000,1.00000}%
\definecolor{mycolor2}{rgb}{1.00000,0.80000,0.00000}%
\begin{tikzpicture}

\begin{axis}[%
width=1.8in,
height=1.4in,
at={(1.752in,0.688in)},
scale only axis,
log origin=infty,
xmin=0.4,
xmax=3.4,
xtick={1,2,3},
xticklabels={{Mice},{Elephant-128MB},{Elephant-64MB}},
tick label style={font=\tiny},
ymin=50,
ymax=110,
ylabel style={font=\tiny},
ylabel={Perecentage (\%)},
axis background/.style={fill=white},
xmajorgrids,
ymajorgrids,
legend style={at={(0.5,1.03)},line width=0.5pt, font=\tiny, anchor=south, legend columns=3, legend cell align=left, align=left, draw=white!15!black}
]
\addplot[ybar, bar shift=-.6cm, fill=mycolor1, draw=black, area legend] table[row sep=crcr] {%
1	100\\
2	100\\
3	100\\
};
\addlegendentry{In-network}

\addplot[ybar,bar shift=-.2cm, fill=mycolor2, draw=black, area legend] table[row sep=crcr] {%
1	100\\
2	100\\
3	100\\
};
\addlegendentry{Centralized}

\addplot[ybar, bar shift=.2cm, fill=red, draw=black, area legend] table[row sep=crcr] {%
1	100\\
2	98.2\\
3	81\\
};
\addlegendentry{OpenSample-TCP(1K)}

\end{axis}
\end{tikzpicture}%
        \caption{500 Flows}
        \label{fig:me1000}
    \end{subfigure}
    \caption{Accuracy comparison of FD schemes in different traffic configurations during the pure scenario.}\label{fig:pure}
\end{figure*}

\subsection{Detection Accuracy}\label{sec:preresults}
The FD mechanisms may have true-negatives (i.e., an EF is considered to be an MF) and false-positives (i.e., considering an MF as an EF). In particular, the percentage of true-negative incidents is critical as keeping the EF on the path of MFs can cause a sever queuing delay for delay-sensitive flows. To evaluate the detection accuracy, both of the FD schemes is investigated under different flow scenarios; pure MFs, pure EFs, as well as a mix of MFs and EFs. Fig.~\ref{fig:accuracy} illustrates the accuracy levels of different schemes when 100, 500, and 1000 flows are exchanged. Both of the FD schemes obviously achieve a 100\% detection accuracy, i.e., the percentage of true-negative incidents is 0\%, in all scenarios. The high accuracy of the proposed FD mechanisms is mainly because of the fact that we only need to subtract the sequence number of the ACK messages from the initial sequence number. Since existing solutions' accuracy varies for different threshold values, we compare our methods with the best case performance of the existing solutions. Fig.~\ref{fig:accuracy} clearly shows that sFlow provides quite a low accuracy in comparison with LightFDG. However, OpenSample-TCP obtains around 90\% accuracy for a varying number of flows. Therefore, we consider only the OpenSample-TCP for comparison in the remainder of the evaluation.

Fig. \ref{fig:pure} depicts the performance comparison of various schemes for 10, 100, and 500 flows in pure MF and EF scenarios\footnote{During the performance evaluations, Mininet is observed to crash in 1000 EF case. Since such errors have serious impacts on the evaluation accuracy, we focus on the comparisons between 10, 100 and 500 flows.}. Similar to the mixed scenarios, LightFDG schemes achieved a high level of accuracy in both pure MF and pure EF cases. Also, OpenSample-TCP attained 100\% in pure MF evaluation, and 90\%-98\% in EF experiment. The accuracy increases with the increase in traffic load due to the long transmission delay of EFs during high congestion. In order to examine both schemes in a short transmission delay and high traffic loads, flow size is reduced to 64MB. Performing the 10, 100, and 500 experiments on the OpenSample-TCP algorithm, a high percentage of true-negative incidents is observed; 40\%, 35.3\%, and 19\% respectively, which in turn is corresponding to 60\%, 64.7\%, and 81\% accuracy. On the same evaluation basis, the proposed FD solutions achieved 0\% of true-negative incidents. 

At this point, it is necessary to discuss the reasons behind the low performance of sampling based methods, which is summarized as follows: 
\begin{itemize}
\item In sFlow, packet sampling randomly captures one out of every $S$ packets and forward it to a collector for analysis, where $S$ is the sampling rate. Accordingly, the collector estimates the flow size by multiplying $S$ with the size of the collected packets from the same flow. However, the calculated flow size is practically less than the actual one especially in bursty traffic scenarios where a large number of packets arrives in a time less than the sampling period. Therefore, detection speed of the sampling methods heavily depends on the flow size estimation as well as the sampling and flow arrival rates
\item  A single TCP packet has partial insight about a flow class. To make accurate and timely decisions, random sampling methods are required to collect multiple samples from the same flow as soon as possible, which is hard to ensure if flows arrive and complete much faster than the sampling rate. 
\item Random sampling has a low efficiency since the captured samples could be from the already classified flows or a flow completed before it has been classified.
\end{itemize}
Although these drawbacks can be partially mitigated by sampling at higher rates, resulting network overhead can exhaust the processing power of switches. Accordingly, the 1-to-1000 sampling rate is widely selected in literature~\cite{DevoFlow, plank, opensample} to strike a good trade-off between the detection performance and overhead. In Fig.~\ref{fig:accuracy}, sFlow performance is enhanced by increasing the sampling rate from 100 to 500. However, increasing the sampling rate is always not a solution since increasing from 500 to 1000 did not deliver a significant improvement. 

On the other hand, OpenSample-TCP is built on top of sFlow to provide a better performance by 1) using TCP packet headers to enhance the efficiency of sFlow and 2) setting a shorter control loop period (100 ms) than that of sFlow (1-5 s). However, OpenSample-TCP still needs high sampling rates as stated in \cite{opensample} to learn accurate and up-to-date statistics about the link utilizations. Notice that OpenSample-TCP requires several packets from the same flow to measure the size by comparing the EF threshold with difference of the first and last captured packets. Therefore, both OpenSample-TCP and sFlow show low accuracy in case of small EF sizes (e.g., 12.8 MB and 32 MB ~\cite{dctcp, vl2}) because it is not possible to collect enough number of samples during the EF flow lifetime.

To this end, a high accuracy is intuitively expected from the proposed solutions since both centralized and in-network schemes are configured to capture almost every ACK. To this end, detection speed must be questioned for such high accuracy performance. In addition to its decentralized nature, the location of the in-network scheme in the pipeline of the packet paths makes it highly accurate and fast. Nonetheless, centralized scheme speed is enhanced by 1) steering the ACK packets to a separate virtual interface, 2) classifying the flows of some applications based on the main flow information, and 3) stop capturing ACK messages of the flows which are already classified.

\subsection{Detection Speed}
\begin{figure}[t!]
    \centering
    \begin{subfigure}[b]{0.48\columnwidth}
\includegraphics[width=1 \columnwidth]{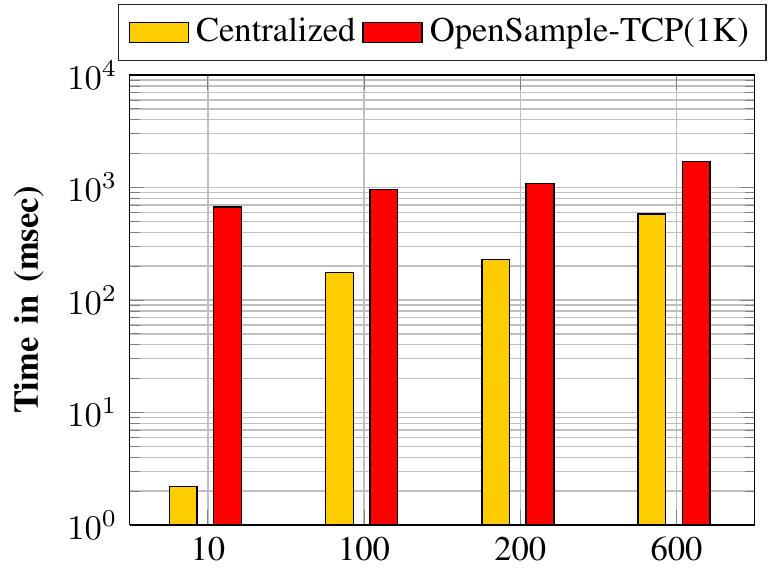}
        \caption{Centralized vs. Open Sample}
\label{innetworkspeed}
    \end{subfigure}
    \begin{subfigure}[b]{0.48\columnwidth}
 \includegraphics[width=1\columnwidth]{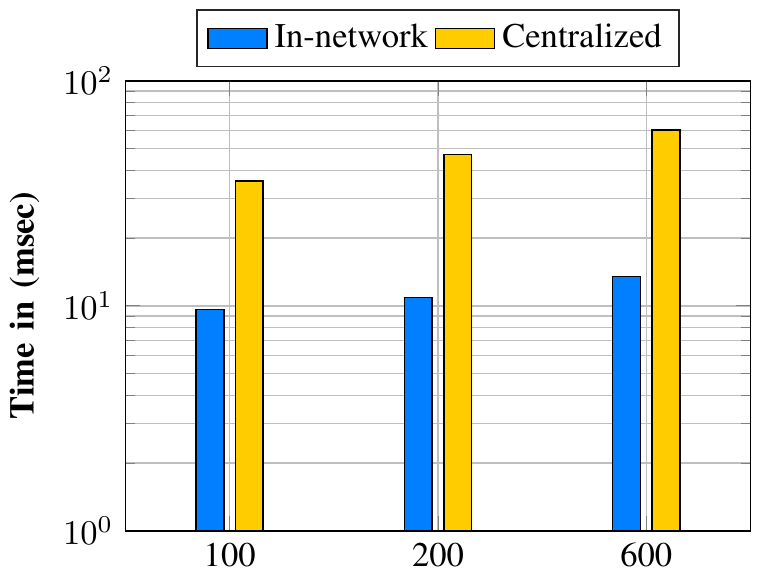}
        \caption{Centralized vs. in-network}
\label{fig:networkandcentral}
    \end{subfigure}
    \caption{Detection speed comparisons under various link loads.}
    \label{fig:speed}
\end{figure}

In order to investigate the detection speeds, we examine different scenarios of network loads; 10, 100, 200, or 600 flows. Notice that these are not the total number of flows in the evaluated DCN, they are rather the number of flows that have been steered into a single link where the collector is implemented. The average speeds of detection during the different traffic loads and configurations are illustrated in Fig.~\ref{innetworkspeed}. According to the load on the path, which range from 2.2 to 580 milliseconds, the centralized scheme is 223.67$\times$ to 3$\times$ faster than OpenSample-TCP. Such performance degradation is because of that EF detection is slower than the measured link utilization. That is, OpenSample-TCP~\cite{opensample} can measure the link utilization in 100 milliseconds while it needs about 1 second to detect EFs in this evaluation. The comparison between centralized and in-network scheme is depicted in Fig.~\ref{fig:networkandcentral} where in-network scheme provides a faster performance especially if link loads increase. This is expected since having large number of flows cause more overhead in the CU.

\begin{figure}[t!]
    \centering
    \begin{subfigure}[b]{0.48\columnwidth}
       \includegraphics[width=\textwidth]{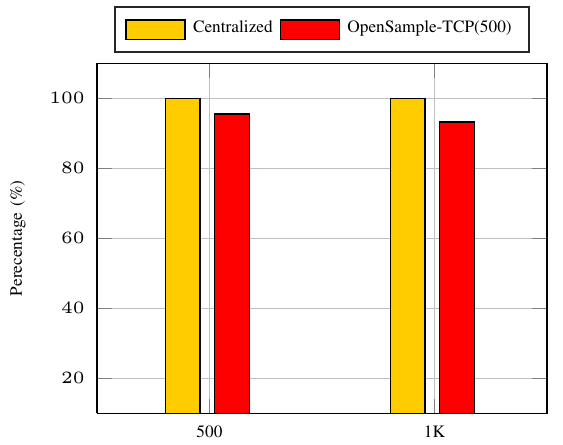}
        \caption{Accuracy}
\label{newaccuracy}
    \end{subfigure}
    \begin{subfigure}[b]{0.48\columnwidth}
       \includegraphics[width=\textwidth]{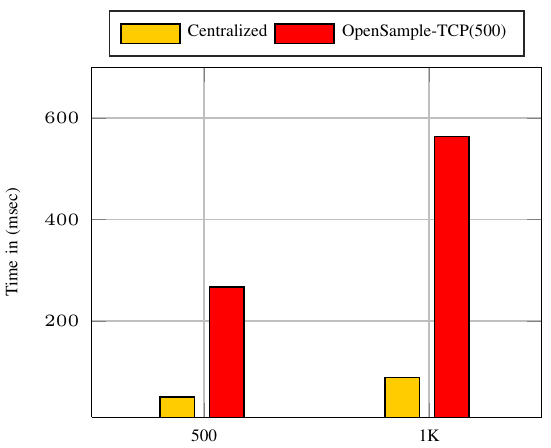}
        \caption{Speed}
\label{newspeed}
    \end{subfigure}
    \label{newresu}
    \caption{Speed and accuracy comparisons under varying flow sizes.}\label{fig:speed_new}
\end{figure}

A varying flow size setup is also investigated as follows: Every MF (EF) randomly selects its size from a range of 10KB to 1MB (10MB to 128MB), where threshold is set to 1MB. Under 500 and 1K traffic loads, we consider 1-to-500 and 1-to-100 sampling rates for OpenSample-TCP and sFlow. Since the in-network scheme is already shown to perform better, we narrow our focus on the comparison with the centralized scheme. For 1-to-500 sampling rate, Fig.~\ref{newaccuracy} shows that the centralized scheme still achieves 100\% accuracy while OpenSample-TCP at 500 load reaches about 95\%, which reduces slightly at 1K load. The Fig.~\ref{newspeed} demonstrates the superiority of the centralized scheme's speed compared to OpenSample-TCP at 500 load. On the other hand, setting sampling rate to 1-to-100 results in a chain of CU failures and 100\% CPU utilization, hence, a comparison between LightFDG and OpenSample schemes was not possible. Although sFlow did not encounter such technical problems since it uses a different collector, both 1-to-100 and 1-to-500 scenarios detected more than 50\% of the flows as EF. That is, more than 55\% of the MFs are considered EF. Hence, we omitted the sFlow results and content ourselves with the comparison of the centralized and OpenSample-TCP schemes.

\begin{table*}[t]
\scriptsize
\captionsetup{justification=centering}
\caption{\\The arrival-rate of LightFDG ACK messages. } % title of Table
\centering % used for centering table
%\resizebox{\columnwidth}{!}{
\begin{tabular}{c c c c c c c c} % centered columns (4 columns)
\hline\hline \\[0.01ex]%inserts double horizontal lines
Dataset(ID) & Dataset ACK/Sec & LightFDG ACK/Sec & Ratio & Dataset(ID) & Dataset ACK/Sec & LightFDG ACK/Sec & Ratio\\ [0.5ex]
\hline \\ [0.001ex]% inserts single horizontal line
Inter. 1 & 3854.2 & 62.6 & 61.6x & Univ. 2.0 & 5.1 & 3.33 & 1.5x \\ [1ex]
Inter. 2 & 3615 & 62.5 & 57.84x & Univ. 2.1 & 11.9 & 0.798 & 15.1x \\ [1ex]
Inter. 3 & 137.5 & 4.15 & 33.13x & Univ. 2.3 & 6.892 & 0.19 & 36.3x\\[1ex]
\hline %inserts single line
\end{tabular}
%}
\label{table:LightFD_ack} % is used to refer this table in the text
\end{table*}

\subsection{Detection Overhead}
In this section, we measure the FD overhead on the CU and the network. Even though the previous network setup and topologies are still valid, the CU is installed in a virtual machine with four processors and 10GB memory. To collect the statistics (e.g., CPU and memory utilization) of the CU, we use a \texttt{psutil} python library~\cite{psutil}, configure \texttt{top} Linux tool to collect the device statistics for every second, and store them on a particular file. For all evaluations, the CU needs to set up the network and prepare the OvS switches with necessary configurations as well as the sFlow rules. This start-up phase consumes about 100\% of a single CPU processing capacity and installs about 7KB of flow-entries. The communication and computation overhead of the in-network scheme is lower than that of the centralized scheme. However, the computation of in-network scheme is at the servers, while the CU is responsible for executing the centralized algorithm. It is also observed that the computation overhead of the in-network FD scheme is very small.

 \begin{figure}[t]
 \centering
  \includegraphics[width=0.8\linewidth, height=2in]{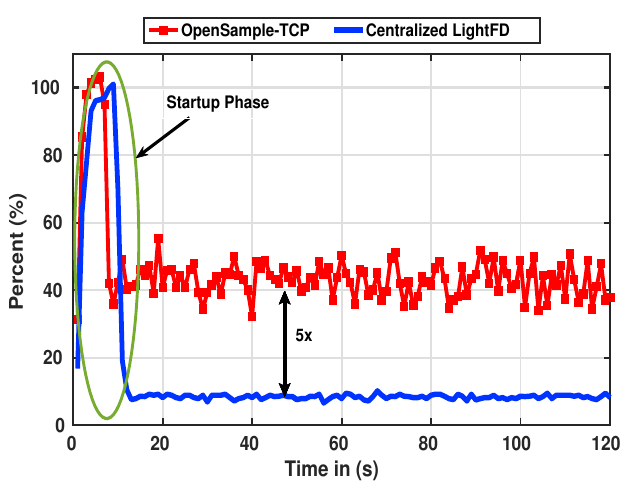}
\caption{The processing overhead during the first 120 sec.}
\label{visperf}
\end{figure}

\begin{figure*}[hbtp!]
  \includegraphics[width=1\linewidth, height=2in]{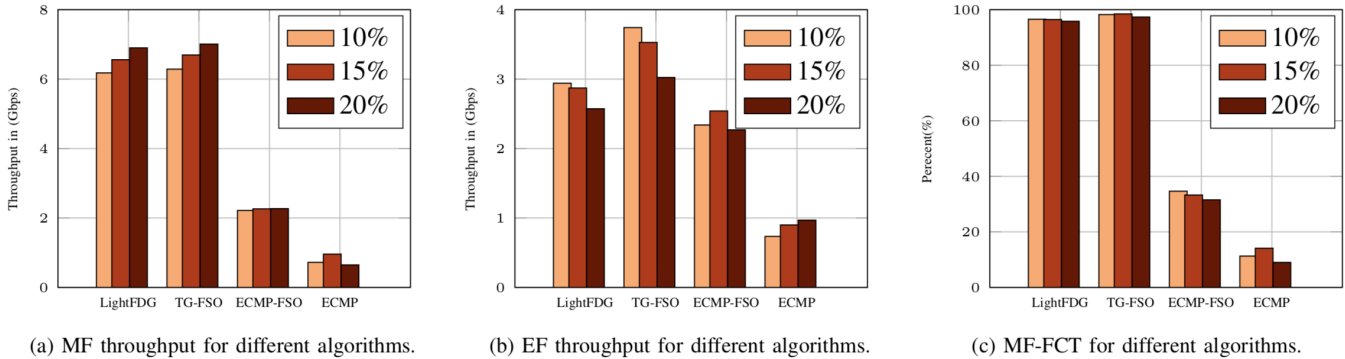}
    \caption{Throughput and FCT performance comparisons during different traffic scenarios.}\label{fig:thr}
    \label{fig:thr}
\end{figure*}

Since the centralized FD configures sFlow in the virtual interfaces with a fast sampling rate, the communication and the processing overhead of the centralized FD is expected to be worse than the others. The OpenSample-TCP continues the collection of samples for all of the flows regardless whether the flow has been classified or not. Thereby, the communication overhead and the processing capacity have not been improved after the classification. Hence, the total communication overhead has increased with the increase in the size of the transmitted file. On the contrary, centralized FD has the overhead minimization techniques that capture only the ACK packets of the unclassified flows. This is justified when the flow size is increased. We found that the communication overhead of the centralized FD is about 70KB while it is 63KB for OpenSample-TCP. For larger flow sizes (e.g., 256MB), the communication overhead of the centralized FD interestingly remains close to 70KB while it increases dramatically for OpenSample-TCP.

Thanks to its simplicity and faster detection, the computational overhead of the centralized FD is 5$\times$ lower than that of the OpenSample-TCP, and close to that of sFlow (1-to-10k). The results are illustrated in Fig.~\ref{visperf} where statistics are collected during the transmission of 1k mix-flows. When we run the proposed overhead minimization algorithm on the enterprise and DCN datasets, the overhead of ACK messages reduced to minimal values compared to the ACK arrival-rate of the datasets themselves. The results are shown in Table.~\ref{table:LightFD_ack}, ranging from 1.5$\times$ up to 61.6$\times$. Moreover, the collected results from \texttt{top} show that the centralized LightFDG and other solutions have a small impact on the memory ($\leq$ 1\%).

\subsection{Network Throughput Results}
Throughout this subsection, the size of MFs and EFs are set to 50KB and 128MB, respectively. Since the links in Mininet are limited by the processing capacity of the host machine, we configured the FSO-links with 10 Gbps and the cabled links with 1 Gbps which means FSO links are ten times faster than wired DCN links. The WDM-FSO links can be thought as parallel independent channels, however, they are coupled by the total available lightpath power intensity. Accordingly, in this evaluation part each FSO link consists of 4 wavelengths, which are realized as virtual links in the simulation. Since the emulator is limited in DCN size, optical channel gains are not distinguishably different due to similar link distances and thus assumed to be identical without loss of generality. 

We use MapReduce to mimic workloads of real DCNs whose shuffle-phase communication pattern has $k$ servers from every rack communicate with another $k$ servers in a different rack. For instance, the hosts in $\mathcal{R}_{i}$ have been divided into two sets and every set has $k$ servers, e.g., 4 MF and one EF. Each of $k$ server is communicating with $k$ servers of rack $j$, $j \neq i$, different than other $k$ servers of the same rack. For accurate results, we evaluated the routing algorithms with various percentage of EF: 10\%, 15\%, and 20\%. We set $k=20$ in all evaluation scenarios, that is, the number of MFs in the 10\% scenario is 20, and the number of EFs is one.

For the network throughput results, we consider the following cases: 
\begin{enumerate}
\item \textit{Equal-Cost-Multi-Path routing (ECMP)~\cite{ecmp}} is a widely used DCN routing method which uses the packet header information, such as the IP/MAC addresses and TCP port numbers, as a key for a hash function. The outgoing path is the output hash value modulo the number of outgoing paths. This strategy splits the flows among available paths. Since the header information for an individual flow is the same during the session, the packets of the same flow are always forwarded via the same path.
\item  \textit{ECMP-FSO} is an ECMP routing algorithm supported by the FSO technology. In this routing method, the link capacity is equally divided between the wavelengths, that is, the capacity of every wavelength is fixed to 2.5 Gbps. Each flow was assigned to a single wavelength. However, when flows are more than the available number of lightpaths, the packets of the waiting flows are enqueued until a lightpath is available for transmission. Since wavelengths capacities are fixed and uniform, this case can be regarded as a DCN with fiber connections. 
\item \textit{FG-FSO} refers to the proposed FGF with a priori knowledge of flow classifications as in \cite{Celik2018Design,Celik2019Design}.
\item  \textit{LightFDG} is the proposed 3-step optical FG algorithm employing the distributed FD mechanism. 
\end{enumerate}
Notice that the first three cases have the perfect and a priori knowledge of flow classifications. By considering the first three cases, we compare the virtues of re-configurable flexibility of WDM-FSO based DCNs and bandwidth utilization performance of FG with wired DCNs, i.e., first two cases. On the other hand, the last case shows the impact of in-network LightFDG scheme on the proposed FG policy

Fig. \ref{fig:thr} compares the throughput and FCT performance of above cases under various work load. Apparently, LightFDG provides much superior performance than the ECMP-FSO by grooming and forwarding each class separately over lightpaths provisioned based on the traffic conditions and QoS demands. Notice that ECMP does not groom traffic and distinguish packets of different flow types while forwarding them over multiple paths for the sake of load balancing. Nonetheless, both MFs and EFs experience performance degradation. In particular, routing MFs along with EFs has a more significant impact on MF throughput (Fig. \ref{fig:thr}.a) and FCT (Fig. \ref{fig:thr}.c).

Since the time constraint is 1ms, the bandwidth demand for 5\%, 10\%, and 15\% scenarios are 7.6Gbps, 7.2Gbps, and 6.8Gbps, respectively. As we initially treat all flows as MFs, EF detection delay has no impact on MF performance as shown in Fig. \ref{fig:thr}.a. That is why LightFDG achieves the same throughput of FG-FSO for MFs.  On the other hand, Fig. \ref{fig:thr}.b shows that the impact of time elapsed for detecting EFs has a distinguishable impact on throughput albeit the fast and accurate FD performance of the in-network scheme. Referring to the detection speed comparisons in previous subsections, detection delays caused by the CU overhead would further deteriorate the system performance. In this regard, OpenSample TCP is expected to severely deteriorate the overall network throughput because of the poor detection speed and accuracy. Finally, Fig.~\ref{fig:thr}.c demonstrates how the LightFDG and FG-FSO achieved about 98\% of the flow completion demand of MFs, while about 70\% of the MFs in ECMP-FSO and exceeded the time constraint and about 88\% of the MFs in ECMP exceeded the time constraint (i.e., complete after the 1 ms time constraint).

\section{Conclusions}
\label{sec:conclusion}
LightFDG is an integrated approach to flow detection and grooming in optical wireless data center networks (DCNs). The LightFDG optically grooms flows of each class into R2R flows, which are separately forwarded over lightpaths of separate virtual topologies dedicated to each flow class. Since high speed and accurate flow-detection mechanisms are necessary to prevent EFs from congesting the MF lightpaths, LightFDG leverage TCP behaviors to develop a generic FD scheme that is suitable for in-network and centralized implementation. Numerical results show that classification accuracy and speed have a significant impact on the FGF performance. Performance evaluations also exhibits that proposed FD mechanisms outperforms sampling based methods in speed, accuracy, and overhead.   
\bibliographystyle{IEEEtran}
\bibliography{Final}

\begin{IEEEbiography}[{\includegraphics[width=1.1in,height=1.25in]{authors/Amer}}]{Amer AlGhadhban}(S'18-M'19) is an assistant professor with the Electrical Engineering Department in the College of Engineering at the University of Hai'l.  He obtained B.S. and M. Sc. degrees in Computer Engineering from King Fahd University of Petroleum and Minerals (KFUPM) in (2012). He received his Ph.D. in Electrical Engineering from King Abdullah University of Science and Technology (KAUST) in (2019). He was a Cisco Academy instructor and during that he earned multiple professional certificates: Cisco Certified Network Associate (CCNA), Cisco wireless, SANS- GIAC Certified Firewall Analyst (GCFW-Gold), Certified Ethical Hacker (CEH), Security Certified Network Engineer (SCNP), and SANS local mentor. His research interests span over multiple fields in networked systems and security.

\end{IEEEbiography}
\vspace*{-2\baselineskip}

\begin{IEEEbiography}[{\includegraphics[width=1.1in,height=1.25in]{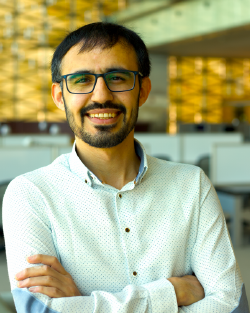}}]{Abdulkadir Celik}(S'14-M'16-SM'19) received a B.S. degree in Electrical-Electronics Engineering from Selcuk University, Konya, Turkey in 2009. He received M.S. degrees in Electrical Engineering in 2013 and in Computer Engineering in 2015, and a Ph.D. degree in co-majors of Electrical Engineering and Computer Engineering in 2016 from Iowa State University, Ames, IA, USA. He is currently a senior post-doctoral fellow in Communication Theory Laboratory at KAUST. His research interests include beyond 5G networks, wireless data centers, Internet of underwater things, and flying networks.
\end{IEEEbiography}
\vspace*{-2\baselineskip}

\begin{IEEEbiography}[{\includegraphics[width=1.1in,height=1.25in]{authors/Shihada}}]{Basem Shihada}(SM'12)  is an associate \& founding professor in the Computer, Electrical and Mathematical Sciences \& Engineering (CEMSE) Division at King Abdullah University of Science and Technology (KAUST). He obtained his PhD in Computer Science from University of Waterloo. In 2009, he was appointed as visiting faculty in the Department of Computer Science, Stanford University. In 2012, he was elevated to the rank of Senior Member of IEEE. His current research covers a range of topics in energy and resource allocation in wired and wireless networks, software defined networking, internet of things, data networks, network security, and cloud/fog computing. 
\end{IEEEbiography}
\vspace*{-2\baselineskip}

\begin{IEEEbiography}[{\includegraphics[width=1.1in,height=1.25in]{authors/Alouini}}]{Mohamed-Slim Alouini} (S'94-M'98-SM'03-F'09) was born in Tunis, Tunisia. He received the Ph.D. degree in Electrical Engineering from the California Institute of Technology (Caltech), Pasadena, CA, USA, in 1998. He served as a faculty member in the University of Minnesota, Minneapolis, MN, USA, then in the Texas A\&M University at Qatar, Education City, Doha, Qatar before joining King Abdullah University of Science and Technology (KAUST), Thuwal, Makkah Province, Saudi Arabia as a Professor of Electrical Engineering in 2009. His current research interests include the modeling, design, and performance analysis of wireless communication systems.
\end{IEEEbiography}

\end{document}